\pdfoutput=1
\documentclass[12pt, draftclsnofoot, onecolumn]{IEEEtran}
\IEEEoverridecommandlockouts

\ifCLASSINFOpdf
\else
\fi
\usepackage{amsmath}
\usepackage{amssymb}
\usepackage{bm}
\usepackage{bbm}
\usepackage{xcolor}
\usepackage{setspace}
\usepackage[ruled,vlined,linesnumbered]{algorithm2e}
\usepackage{setspace}
\usepackage{algpseudocode}
\usepackage{yfonts}
\usepackage{url}
\usepackage{hyperref}
\usepackage{amsthm}
\newtheorem{theorem}{Theorem}%[section]
\newtheorem{definition}{Definition}

\newtheorem{prop}{Proposition}

\newcommand{\SUM}[2]{\overset{#1}{\underset{{#2}}{\sum}}}

\newcommand{\R}{\mathbb{R}}

\newcommand{\algsmall}[1]{\textproc{\small{#1}}}

\usepackage{epsfig}
\usepackage{booktabs,siunitx}
\usepackage{subfloat}
\usepackage{subcaption}
\usepackage{lineno}
\usepackage{scalerel} % Scaling subscripts
\newtheorem{assumption}{Assumption}
\begin{document}
\linespread{1.62}
\title{Multi-task Federated Edge Learning (MtFEEL) in Wireless Networks}

\author{Sawan Singh Mahara, Shruti M., B.~N.~Bharath, Akash Murthy\footnote{The authors are with the Department of electrical engineering at IIT Dharwad, Dharwad, Karnataka. \texttt{email:}\{mshruti32,ssmahara96\}@gmail.com, bharathbn@iitdh.ac.in.\\Akash Murthy is at Euprime Pvt. Ltd. email:akash@euprime.org }}% <-this % stops a space
\maketitle
\begin{abstract}
Federated Learning (FL) has evolved as a promising technique to handle distributed machine learning across edge devices.
A single neural network (NN) that optimises a global objective is generally learned in most work in FL, which could be suboptimal for edge devices.
Although works finding a NN personalised for edge device specific tasks exist, they lack generalisation and/or convergence guarantees.
In this paper, a novel communication efficient FL algorithm for personalised learning in a wireless setting with guarantees is presented. The algorithm relies on finding a ``better`` empirical estimate of losses at each device, using a weighted average of the losses across different devices. 
It is devised from a Probably Approximately Correct (PAC) bound on the true loss in terms of the proposed empirical loss and is bounded by (i) the Rademacher complexity, (ii) the discrepancy,  (iii) and a penalty term. 
Using a signed gradient feedback to find a personalised NN at each device, it is also proven to converge in a Rayleigh flat fading (in the uplink) channel, at a rate of the order $\max\left\{\frac{1}{SNR},\frac{1}{\sqrt{T}}\right\}$.
% max{1/SNR,1/sqrt(T)}
Experimental results show that the proposed algorithm outperforms locally trained devices as well as the conventionally used FedAvg and FedSGD algorithms under practical SNR regimes.

\end{abstract}
\vspace{-6mm}

\IEEEpeerreviewmaketitle
\newpage
\section{Introduction}
% The proliferation of internet services and smartphones have resulted in several interesting applications requiring machine learning (ML) algorithms to work in a distributed fashion (see \cite{kraska2013mlbase}) with heterogeneous devices and data. 
The wide spread adoption of smartphones and internet services with considerable computing capabilities has enabled machine learning (ML) algorithms to work in a distributed fashion (see \cite{kraska2013mlbase}). Since the data and the edge devices are heterogeneous, a new paradigm called \emph{Federated Learning (FL)} (see \cite{konevcny2016federated_ondevice,mohri2019agnostic, bonawitz2019towards,yang2019federated}) has emerged, where the data is distributed while the central node controls the exchange of the data. FL is faced with several challenges such as (i) \emph{stragglers}, where edge devices leave the network and stops contributing to the FL process, without warning, (ii) untimely updates from heterogeneous edge devices with varying computation power, (iii) statistical heterogeneity in data, (iv) privacy concerns, and (v) communication between edge devices and a central node or a Base Station (BS) being expensive \cite{li2018federated}.
% The task of FL can essentially be boiled down to a distributed implementation of the Stochastic Gradient Descent (SGD) algorithm \cite{boyd2003subgradient} with heterogeneous requirements.
%In general, FL aims to implement the Stochastic Gradient Descent (SGD) algorithm in a distributed fashion  \cite{boyd2003subgradient} with heterogeneous requirements.
A de facto version of FL algorithm used in applications such as next word prediction in mobile keyboards is called Federated Averaging (FedAvg) \cite{hard2018federated} which aims to optimise a global objective. It has been shown that FedAvg outperforms models trained using data from individual devices \cite{mcmahan2017federated}. FedAvg involves performing multiple rounds of SGD on a subset of users in the network, and communicating the resulting neural network to a central BS, where they will be averaged and communicated back to devices. This process is repeated until convergence. Although successful, FedAvg does not fully address the issues of data heterogeneity \cite{mcmahan2016communication}, and is known to diverge in non-i.i.d settings due to model drift \cite{li2018federated}. To circumvent this, a modified version of the FedAvg algorithm called \texttt{FedProx} is proposed in \cite{li2018federated}. Subsequently, numerous algorithms such as SCAFFOLD \cite{karimireddy2020scaffold}, MIME \cite{karimireddy2020mime}, SlowMo \cite{wang2019slowmo}, QG-DSGDm \cite{lin2021quasi} have been proposed to handle the model drift problem.
Other extensions include FL algorithms to handle differential privacy \cite{mcmahan2017learning}, secure aggregation \cite{bonawitz2017practical}, and scenarios with less number of devices participating in the FL process \cite{yang2019federated}. In all of the above work (except \cite{li2018federated}), a single model is returned that optimizes a global objective, and hence does not perform well in the case of heterogeneous data. Therefore, there is a pressing need towards designing a personalized FL system in wireless scenarios, with less communication between the BS and edge devices. This problem is addressed in this paper.
\subsection{Related Work and Motivation}
% A naive implementation of the SGD requires repeated exchanges of gradients of the ML model (typically a vector of a million or more entries) leading to huge radio resource requirements. 
A naive implementation of the FL using SGD would require repeated exchanges of gradients of the losses (typically a vector of a million or more entries), which leads to huge radio resource requirements. 
This communication overhead can be overcome by compressing the gradient information before being transmitted \cite{konevcny2016federated}.
A simple way of compressing the gradient is to use the sign of the gradient called SIGNSGD. A detailed theoretical analysis of a majority vote based SIGNSGD with non-convex loss is provided in \cite{bernstein2018signsgd}.
In wireless edge devices, the number of updates to each device can be further reduced by exploiting the nature of the wireless medium leading to a solution called \emph{over-the-air aggregation} also called \emph{over-the-air computation} \cite{zhu2019broadband,amiri2019machine,yang2020federated}. An extension of this called one-bit broadband digital aggregation (OBDA) is proposed in \cite{zhu2020one}.
Further, a similar work with client scheduling and resource block allocation with improper channel state information (CSI) is considered in \cite{jointclientscheduling}. The authors in \cite{samarakoon2018federated} study the impact of wireless channel hostilities under some assumptions on the CSI. Mobile edge computing devices that have resource constraints and spotty wireless communication links pose their own set of challenges. Providing some guarantees on the model accuracy achievable in such situations has been looked at by the authors of \cite{mec}. In all of the above literature, an estimate of the average loss is minimised in a Federated fashion to obtain a single neural network.

The above work fails to provide generalisation guarantees and doesn't work well on device specific tasks.
In non-IID settings, the performance at a device with heterogeneous data distributions is difficult to improve with a single model. This issue can be tackled by learning multiple models for different target distributions.
In this light, a distributed multi-task learning (MTL) algorithm called \textsc{Mocha} tries to learn a single neural network optimised to its task (see \cite{NIPS2017_6211080f}). Here,
% In this light, \cite{NIPS2017_6211080f} proposed \textsc{Mocha} a distributed Multi-Task Learning (MTL) framework that considers each device as a task and learns a single model per device.
each task is learned by solving a primal-dual optimisation problem in a convex setting.
The mixture methods FL framework (see \cite{deng2020adaptive}, \cite{hanzely2020federated}, \cite{mansour2020three}, \cite{li2021ditto}) achieve some personalisation by combing the model parameters obtained by training a local model and a global model.
% and \cite{peterson2019private} ensures that this is done with privacy guarantees.
Global and local parameter mixing can be done across neural network layers, by incorporating the lower layers to adapt to each device's data while having the higher layers shared among other devices (see \cite{liang2020think}).
%Check if this line is fine to delete
Adapting the existing federated averaging algorithm to mixture methods, the authors of \cite{fallah2020personalized} utilise meta learning to personalise a global model to each device.
Alternatively, if the assumption that the non-IID data are partitioned into groups and can be clustered, Clustered FL (\cite{sattler2019clustered} ,\cite{NEURIPS2020_e32cc80b}, \cite{briggs2020federated}, \cite{mansour2020three}) addresses these challenges by grouping devices with similar distributions to improve model accuracies.
% , transfer learning \cite{zhao2018federated} and meta learning (\cite{jiang2019improving}, \cite{chen2018federated}, \cite{li2019differentially}, \cite{fallah2020personalized}, \cite{singhal2021federated}).
Another approach to improve model accuracies via personalisation is by a weighted combination method. For example, \textsc{FedFomo} \cite{zhang2020personalized} uses the information of how much any device could benefit from another device's model.
Many issues in the above work are as follows
% The above fail to address the following issues.
(i) a lack of a personalised model, (ii) poor performance due to inaccurate estimates of loss functions using local data, (iii) a lack of generalisation guarantees and convergence guarantees. This work address all these issues in a systematic manner.
\subsection{Contributions of the paper}
The setup studied in the paper consists of a wireless network of edge devices (like smart phones) connected to a BS with the goal of learning optimal neural network at each of the edge devices to perform some supervised learning tasks such as classification, prediction, regression, to name a few. In particular, an improved estimate of the loss at each device is used to optimize the neural network weights. An improved estimate at each device is obtained by using weighted loss across all the devices. Naturally, the devices with similar data should be given higher weights. Finding these weights and subsequently the neural network tailored towards the task of each device is a challenging problem of multi-task learning that is addressed in the paper. This paper presents a systematic approach to finding the weights backed by theory. In particular, a Probably Approximately Correct (PAC) \cite{guedj2019primer} bound on the performance of the weighted average losses across devices with respect to the true loss is presented. It is shown that the bound depends on (i) Rademacher complexity; a measure of the complexity of the learning task, (ii) discrepancy; a measure of statistical ``closeness" of the data between any two devices, and (iii) a regularization term on the weights. 
Based on the insights provided by the bound, a distributed learning algorithm to find (i) an estimate of the discrepancy, (ii) a device importance weighting metric and (iii) the weights of neural networks, is presented. 
In the absence of wireless abnormalities, within $T$ communication rounds, the algorithm is shown to converge at a rate of $1/\sqrt{T}$. At the end of the training, all devices are provided with a custom neural network, which characterises the multi-task nature of the algorithm.
% Analysing this in a wireless scenario, it is shown to converge in the Rayleigh fading channel, provided that the SNR is reasonably high. In particular for any given SNR, higher rounds of training $T$ isn't guaranteed to yield performance benefits.
In a Rayleigh flat fading channel scenario, the algorithm is shown to converge provided the $SNR$ is reasonably high. In particular, the convergence as a function of $SNR$ and $T$ is shown to be $\mathcal{O}\left( \max \left\{\frac{1}{SNR}, \frac{1}{\sqrt{T}}\right\} \right)$.
In other words, for a fixed $SNR$, higher rounds of training $T$ does not guarantee to yield better convergence performance. 
This system is simulated using python and tensorflow and the proposed algorithm (MtFEEL) was shown to outperform locally trained neural networks as well as existing state-of-the-art federated algorithms.
% The results in an error free channel are extended to the wireless setting and shown that for SNRs of practical interests, the proposed algorithm outperforms classical approaches such as FedAvg, FedSGD and local training.
The simulations were also performed in a wireless setting and shown to outperform classical approaches such as FedAvg, FedSGD and local training in SNRs of practical interests. However, at lower $SNR$s, performance of the MtFEEL algorithm degrades, in which case it is better to use classical federated algorithms, as indicated by our convergence results. These insights can prove useful in designing next generation wireless networks.
% \subsection{Related Work}
% Classical federated learning scenarios \cite{mcmahan2017communication} consider minimising the average loss of all devices by equally weighting everyone, which is unfair without considering the differing data distributions of each of the devices. The authors of \cite{mohri2019agnostic} tackle the issue of fairness by optimising for the worst possible scenarios, like if devices leave the network prematurely. They however do not take into account the data distributions between devices either. Multi-task learning enables any device to learn a custom hypothesis for itself. The work in \cite{shui2019principled} considers multi-task learning and shows that the exploitation of shared knowledge improves the performance of these individual tasks learned. This proposed work attempts to exploit shared knowledge with the help of a discrepancy metric and a learning guarantee. It also proposes a novel algorithm that takes into account bandwidth constraints and proposes a signed gradient communication, similar to that done in \cite{bernstein2018signsgd}.
% \subsection{Contributions}
The paper is organized as follows. In Sec.~\ref{sec:sys_model}, the system model and the problem considered in the paper are presented. Section \ref{sec:mainresult1} presents the first main result based off of which, the algorithm in the noiseless scenario is presented in Sec.~\ref{Algo_Section}. The convergence results of this algorithm in an ideal, error free regime is presented in Sec.~\ref{sec:convergence_analysis}. Section \ref{sec:rayleigh} extends the convergence result to noisy channel. The performance of the proposed algorithm on real data set is presented in Sec.~\ref{sec:expresults}. Finally, the paper is concluded in Sec.~\ref{sec:conclusion}.
\section{System Model}
\label{sec:sys_model}
\begin{figure}
    \centering
    \includegraphics[scale=0.15]{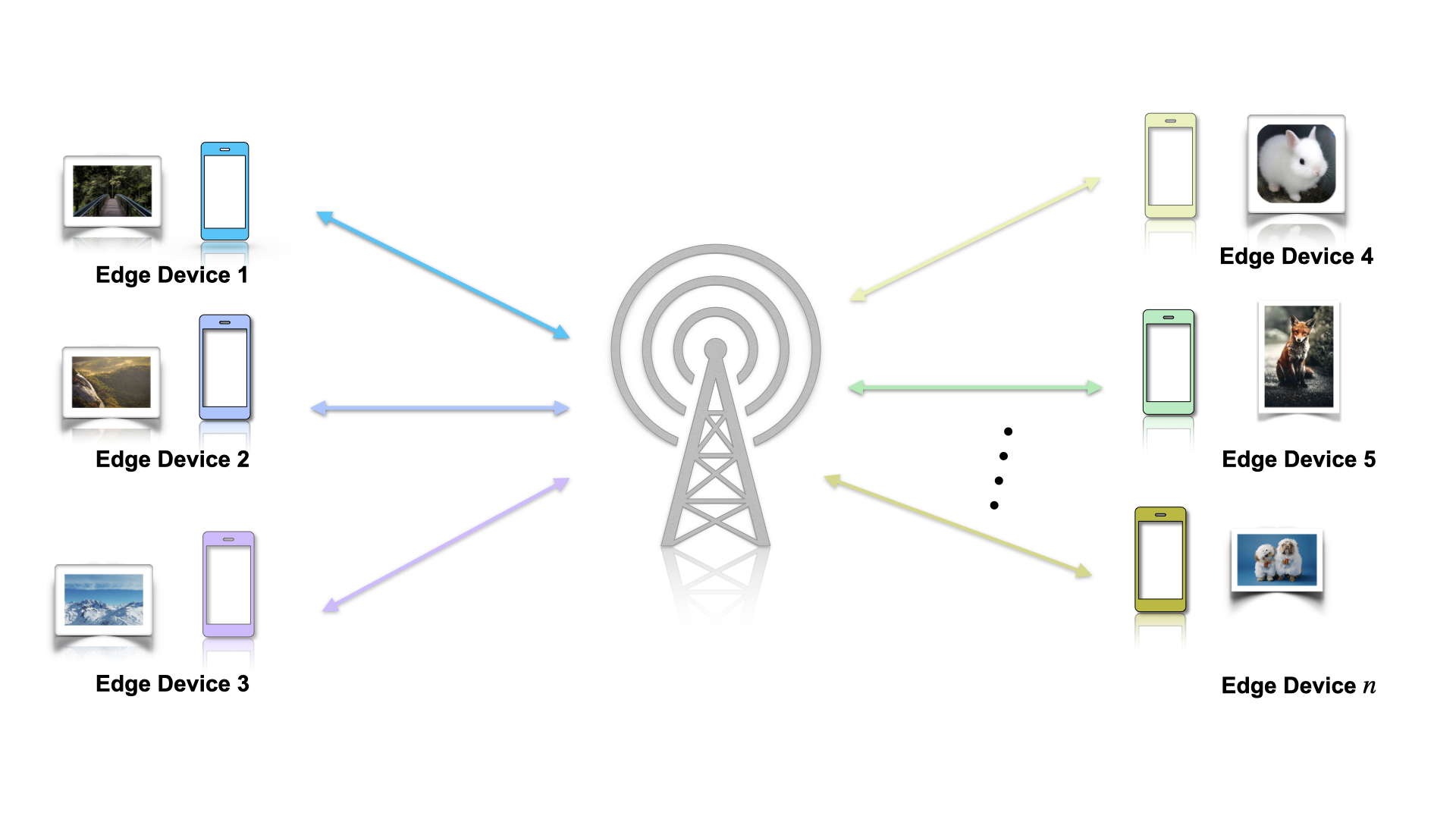}
%\vspace{-0.6cm}
\caption{Federated learning scenario}  
\label{fig:fed_model}
\end{figure}

The paper considers the problem of federated multi-task learning with $N$ devices (example, mobiles) and a BS, as shown in Fig.~\ref{fig:fed_model}.
Each device has a certain task like next word prediction, and the user of the device provides data (supervised) to learn the ML model. Assume that the $k$-th user has $n_k$ training data denoted by $S_k=\{(\bm x_{k1},y_{k1}),(\bm x_{k2},y_{k2}),\ldots,(\bm x_{kn_k},y_{kn_k})\}$, where $\bm x_{ij} \in \mathcal{X}$ is the feature vector corresponding to the $j$-th training example at the $i$-th edge device, and $y_{ij} \in \mathcal{Y}$ is the corresponding label. Let $\bm S:=\{ S_{1}, S_{2}, \ldots S_{N} \}$ be the set of all samples present.  Here, $\mathcal{X} \subseteq \mathbb{R}^d$ and $\mathcal{Y}$ represent the feature space and the output (label) space, respectively. The data at the $k$-th device is drawn in an i.i.d. fashion with distribution denoted by $\mathcal{D}_k$, $k=1,2,\ldots,N$. Further, the data across devices is assumed to be independent but not necessarily identical distributed.\footnote{This assumption is made for the sake of clarity of presentation.} Typically, the devices would not like to communicate the raw samples due to privacy concerns. Therefore, the learning should happen in a federated fashion. The learning rule/hypothesis considered is of the form $h_{\bm{w}_k}: \mathcal{X} \rightarrow \mathcal{Y}$, $k = 1,2, \ldots, N$.\footnote{Note that instead of $\mathcal{Y}$, one can consider $\Delta_y$, the simplex over $\mathcal{Y}$ as well.} It is important to note that any neural network architecture can be characterised in this way. Let $W=\{\bm w_1\ldots \bm w_N\} \in \mathbb{R}^{d \times N}$. Given a feature vector $\bm{x}_k \in \mathcal{X}$, and the corresponding label $y_k \in \mathcal{Y}$, $k=1,2,\ldots,N$, the performance of the neural network at each device $k$ is measured using a loss function $l: \mathcal{Y} \times \mathcal{Y} \rightarrow \mathbb{R}^+$. For the ease of notation, let $\mathcal{L}_k(\bm w) := \underset{(\bm x, y) \sim \mathcal{D}_k}{\mathbb{E}}l\left(h_{\bm w}(\bm x), y\right) $ and the corresponding estimate by $\hat{\mathcal{L}}(\bm w, S_{k}) := \frac{1}{n_k} \sum_{i=1}^{n_k} l\left(h_{\bm w}(\bm x_{ki}), y_{ki}\right)$. Note that the estimate is unbiased, i.e., $\mathcal{L}_k(\bm w) = \underset{S_k \sim \mathcal{D}_k}{\mathbb{E}}\hat{\mathcal{L}}(\bm w, S_{k})$ for any $\bm w \in \mathbb{R}^d$. In the classical federated setting, the goal is to solve the following optimization problem, i.e., find one neural network $\bm w \in \mathbb{R}^d$ across all the devices: 
\begin{equation} 
\label{eq:typical_fl}
\min_{\bm w \in W} \left\{ \Phi_{\bm w} := \sum_{k=1}^{N}\frac{n_k}{n} \mathcal{L}_k(\bm w)\right\}, 
\end{equation}
where the total number of samples $n:=\sum_{k=1}^N n_k$. Note that in practice, the solution to the above can be obtained by using an estimate of the gradient of $ \mathcal{L}_k(\bm w)$ in the SGD algorithm. The challenge in the FL setting is that the overall gradient is the sum of the gradients of individual average losses. A number of solutions such as FedAvg \cite{mcmahan2016communication}, one bit gradient based majority vote \cite{bernstein2018signsgd}, and many more have been proposed in the literature \cite{konevcny2016federated}
% A number of solutions such as FedAvg \cite{mcmahan2016communication}, one bit gradient based majority vote \cite{bernstein2018signsgd}, and many more have been proposed in the literature \cite{konevcny2016federated, mcmahan2016federated}.
%\textcolor{red}{more papers}. 
% \begin{figure}
%     \centering
%     \includegraphics[width=5.2 cm,height = 4.6 cm,keepaspectratio]{Block_Diagram_federated .png}
% \caption{Figure shows a typical FL scenario with next word prediction application.}  \label{fig:fed_model1}
% \end{figure} 
The co-efficient $n_k/n$ will determine the importance of the loss corresponding to the user $k$. However, if the number of samples in the future from user $k$ is less, then the model returned by solving the above problem may result in poor performance across several devices. A solution to the above is to look at the worst case scenario as described below \cite{mohri2019agnostic}:
\vspace{-0.2cm}
\begin{equation} \label{eq:agnostic_problem}
\min_{\bm w \in W} \sup_{\bm{\lambda} \in \bm \Lambda} \left\{ \Phi_{\bm w,\bm{\lambda}} := \sum_{k=1}^{N}\lambda_k \mathcal{L}_k(\bm w)\right\},
\end{equation}
where $\bm{\lambda} := (\lambda_1,\lambda_2,\ldots,\lambda_N)$, and the constraint set $\bm \Lambda \subseteq \Delta_N$ incorporates the prior knowledge on devices that may drop off from the network (see \cite{mohri2019agnostic}). Note that the above problem uses the same neural network $\bm w$ across all the devices, unlike the scheme proposed in this work. The neural network can be made more personalised by using different neural network weights, as done by the authors of \cite{smith2017federated}. This work however, explicitly considers the statistical heterogeneity across devices as well as weighting every loss metric in accordance with the heterogeneity, which leads to solving the following optimization problem. 
\begin{equation} \label{eq:agnostic_problem_diff_weights}
\min_{\bm w_1,\bm w_2, \ldots, \bm w_N} \sup_{\bm{\lambda} \in \bm \Lambda} \left\{ \Phi_{W,\bm{\lambda}} := \sum_{k=1}^{N}\lambda_k \mathcal{L}_k(\bm w_k)\right\}, 
\end{equation}
In order to solve the above problem, the devices should compute an estimate of $\mathcal{L}_k(\bm w_k)$ denoted by
$\hat{\mathcal{L}}(\bm w_k, S_{k}) := \frac{1}{n_k} \sum_{i=1}^{n_k} l\left(h_{\bm w_k}(\bm x_{ki}), y_{ki}\right)$, which can be used as a proxy in \eqref{eq:agnostic_problem_diff_weights}.
Note that the estimate of the average loss of the $k$-th device depends only on its data. However, it is natural to include neighbors' empirical estimate of the average losses while estimating the average loss of the $k$-th device if the neighboring data distribution is ``close" to the distribution of the data of the device. An extreme scenario is that of an i.i.d. data across devices where a simple averaging works well. One approach is to take the average of the empirical loss across all the devices. This may lead to a bad estimate of the average loss since the neighboring data are given equal weights. A way around this problem is to ``optimally" allocate weights across users data. This leads to the following optimization problem that needs to be solved in a federated manner
\vspace{-3mm}
\begin{equation} \label{eq:obj_weighted_hat}
 \min_{\bm w_1, \bm w_2, \ldots, \bm w_N} \sup_{\bm{\lambda} \in \bm \Lambda} \left\{ \hat{\Phi}_{W,\bm{\lambda},\bm{\alpha}}(\bm S) := \sum_{k=1}^{N} \lambda_k \sum_{j=1}^{N} \alpha_{kj} \hat{\mathcal{L}}(\bm w_k, S_j)  \right\}.
\end{equation}
{It is natural to constraint $\alpha_{kj}$ as $\sum_{j=1}^N \alpha_{kj} = 1$}. In the above, let the weights be denoted by $\bm \alpha : = \{\bm \alpha_1,\bm \alpha_2,\ldots, \bm \alpha_N\}$ and $ \bm \alpha_m : =\{ \alpha_{m1}, \alpha_{m2}, \ldots, \alpha_{mN} \}$ for all $m=1,2,\ldots, N$. The neural network weights computed using \eqref{eq:obj_weighted_hat} will be close to the one computed in \eqref{eq:agnostic_problem} provided the gap between $\hat{\Phi}_{W,\bm{\lambda},\bm{\alpha}}(\bm S)$ and ${\Phi}_{W,\bm{\lambda}}$ is small. It is now evident that the federated learning requires users to communicate with the BS and vice-versa. In the following subsection, the channel model for the BS and users to communicate is presented.
\vspace{-5mm}
\subsection{Channel Model}
\label{subsec:channel_model}
The communication model consists of a single BS Single Input Single Output (SISO) system with multiple users or edge devices. In particular, the time is assumed to be slotted, and the channel between any device $k$ and the BS is assumed to be a wireless Rayleigh flat fading channel. The complex baseband received signal $y(t) \in  \mathcal{C}$ at the BS for an input $x_k(t) \in  \mathcal{C}$ by a device $k$ is given by
\vspace{-7mm}
\begin{align}
  y_k(t)= h_k(t) x_k(t) + z_k(t), ~t =1,2,\ldots, \text{ and } k = 1,2,\ldots, N, 
\end{align}
where $h_k(t) \stackrel{\text{i.i.d.}}{\sim} \mathcal{CN}(0,1)$ is the fading channel coefficient between device $k$ and the BS. The noise $z_k(t) \; {\sim} \; \mathcal{CN}(0, \sigma^{2})$ is a circularly symmetric complex Gaussian random variable. Typically, the devices are mobile phones, and hence are power limited. Therefore, the power constraint at the device $k$ is given by $\mathbb{E}{|x_k(t)|^2} \leq P_k$. On the other hand, the BS is assumed to have enough power to communicate without any errors. This assumption is made for the sake of simplicity and the ease of exposition. At any given time slot, it is assumed that the scheduler will assign the channel of bandwidth $B$ to any edge device that wishes to communicate. For example, in an OFDMA system, a resource block is allocated to a user who wishes to transmit. For the sake of simplicity, the impact of the scheduling scheme on the convergence of the proposed algorithm is ignored.    %The analysis of the proposed algorithms in this channel is introduced in section \ref{fading channel} after the initial assumption of an error free communication link, with no distortion i.e., $h_k(t) = 1$ and $z_k(t) = 0 \; \forall \; k $ and $ \forall \; t$. 
In order to device an algorithm, and prove convergence under a noisy communication channel, first a noiseless scenario is considered. Subsequently, the results are extended to study the impact of wireless communication channels on the convergence of the algorithm. The first main result of this paper is to prove a PAC bound  \cite{guedj2019primer} on $\hat{\Phi}_{W,\bm{\lambda},\bm{\alpha}}(\bm S) - \Phi_{W,\bm{\lambda}}$ when the channels between devices and BS are ideal, i.e., without any error. This PAC bound will be used to device a distributed federated algorithm that is shown to outperform state-of-the-art federated algorithms. Subsequently, a convergence guarantee on the proposed algorithm is also provided. The impact of the wireless channel model explained above on the convergence is provided in Sec.~\ref{sec:rayleigh}. The following section presents the main result. 

\section{Main Result - I} \label{sec:mainresult1}
To state the first main result of the paper, the following three quantities will be required (i) Rademacher complexity, (ii) Minimum $\epsilon$-cover, and (iii) discrepancy, which are defined below. 

\begin{definition} 
\label{rademach}
(\textbf{Minimax weighted Rademacher complexity \cite{mohri2019agnostic}})  The Rademacher complexity for the class of neural networks $W$ for a given $\bm{\lambda} \in \bm \Lambda$ is defined as 
\[ 
  \mathcal R_{\bm \lambda}\left(W \right) :=  \underset{\bm S,\bm \sigma}{\mathbb{E}}\left[\sup_{\substack{\bm w_1 \bm w_2, \ldots \bm w_N\\\alpha\in\Delta_N}}\sum_{k,j=1}^{N}  \frac{\lambda_{k} \alpha_{kj}}{n_{j}} \sum_{i=1}^{n_{j}} \sigma_{kj,i}  l \left(h_{\bm w_k}\left(\bm x_{ji}\right), y_{ji}\right)\right],
\]
where the Rademacher random variables $\sigma_{kj,i} \in \{1,-1\}$ for $ k,j = 1, 2, \ldots, N$ occur with equal probability. The max weighted Rademacher complexity is defined as ${\mathcal R_{\bm \Lambda}\left(W \right)= \max_{\bm \lambda \in \bm \Lambda} \mathcal R_{\bm \lambda}\left(W \right).}$
\end{definition}
%Note that the term $\frac{1}{n_{j}} \sum_{i=1}^{n_{j}} \sigma_{kj,i}  l \left(h_{\bm w_k}\left(\bm x_{ji}\right), y_{ji}\right)$ corresponds to the inner product of ranphirical average of the losses at node $j$. If the loss corresponds to $0-1$ loss, then this terms provides the 
This complexity is a measure of how well any members of a class of real valued hypotheses can approximate random noise. The expressibility of the neural network is measured based on how well the hypothesis (model) class fits the noise. The class can therefore be expected to learn more intricate decision boundaries. In order to find the $\sup$ over $\bm \lambda \in \bm \Lambda$, it is useful to quantize the set $\bm \Lambda$ for which the following definition of minimum $\epsilon$-cover comes in handy \cite{40597}.

\begin{definition}
\textbf{(Minimum $\epsilon$-cover \cite{40597})}
\label{min_epsilon_cover}
The set $\left\{\bm v_{1}, \ldots, \bm v_{p}\right\}$ is said to be an $\epsilon$-cover of $\bm \Lambda$ with respect to $\ell_1$-distance if $\bm \Lambda \subseteq \cup_{i=1}^{p} B\left(\bm v_{i}, \epsilon\right)$, where the $L_1$ ball is defined as $B(\bm v_{i}, \epsilon): = \{\bm z \in \bm \Lambda : \Vert \bm z-\bm v_i \Vert_1 < \epsilon \}$. The minimum $\epsilon$-cover $\bm \Lambda_{\epsilon}$ of a set $\bm{\Lambda}$ is any $\epsilon$-cover with the smallest $p$. 

\end{definition}
It is expected that the optimal weights $\alpha_{kj}$ will depend on ``closeness" of the distributions of data across the devices. The following definition provides a measure of the difference in two distributions with respect to a loss function.

\begin{definition}
(\textbf{Discrepancy \cite{mohri2019agnostic}}) Given two data distributions $\mathcal{D}_{k}$ and $\mathcal{D}_{j}$ of the devices $k$ and $j$ respectively, the corresponding discrepancy with respect to the loss $l : \mathcal{Y} \times \mathcal{Y} \rightarrow \mathbb{R}^+$ is defined as $d_{kj}:= \sup_{\bm{w} \in \mathbb{R}^d} \Delta_{kj}(\bm{w})$, where $\Delta_{kj}(\bm{w}) := \left| \mathcal{L}_k(\bm w)- \mathcal{L}_j(\bm w)\right|$.
\end{definition}
Recall that $\mathcal{L}_k(\bm w) := \underset{(\bm x, y) \sim \mathcal{D}_k}{\mathbb{E}}l\left(h_{\bm w}(\bm x), y\right)$.
The first term above corresponds to the average loss at the $k$-th device while the second term corresponds to the average loss at the $j$-th device.
The difference provides a measure of how different the two data distributions are with respect to the neural network $\bm w$.
Maximising over $\bm w \in \mathbb{R}^d$ gives the worst case difference.
If the data at the two devices are i.i.d., then, it is easy to see that the discrepancy is zero.
On the other hand, if the distributions of data at the two devices are different, and the same neural network is not able to classify the data, then the discrepancy is high.
A naive way of using this while devising an algorithm is to allocate higher weights $\alpha_{kj}$ if $d_{kj}$ is small.
In the following, a theoretical result that provides insights on how to choose these weights in a systematic way is provided.
In particular, using the definitions above, a PAC bound on the difference between the true average loss in \eqref{eq:agnostic_problem_diff_weights} and its estimate in \eqref{eq:obj_weighted_hat} is provided in the following theorem.
The proof is relegated to Appendix [\ref{problem_formulation}]. 

\begin{theorem}
\label{MtFEEL_Theorem}
\textbf{(PAC bound)} Assuming that the loss is bounded, i.e., $l(a,b) \leq M \; \forall \; a, b \in \mathcal{Y}$, for every $\epsilon > 0$, with a probability of at least $ 1-\delta $, $\delta > 0$, the following holds 
\begin{equation}
\begin{aligned}
\label{main_bound}
{\Phi}_{W,\bm{\lambda}} \leq        \hat{\Phi}_{W,\bm{\lambda},\bm{\alpha}}(\bm S) +
2\mathcal R_{\bm \Lambda}\left(W \right) + M \texttt{Pen}(\bm{\lambda}, \bm{\alpha}) + MN \epsilon,
\end{aligned}
\end{equation}
where $\texttt{Pen}(\bm{\lambda}, \bm{\alpha}): = \sqrt{ \frac{N}{2}\sum_{j=1}^{N}\sum_{k=1}^{N}\left(\frac{\lambda_{k}\alpha_{k j}}{ n_{j}}\right)^{2}\log\left(\frac{|\Lambda_{\epsilon}|}{\delta}\right)} + 
\frac{1}{M} \sum_{k,j=1}^{N}\lambda_{k}\alpha_{kj}{d}_{kj}$, $d_{kj}$ is the discrepancy, and $\bm \Lambda_{\epsilon}$ is the minimum $\epsilon$-cover of $\bm \Lambda$.
\end{theorem}
The above guarantee suggests that the neural network weights $\{\bm w_1, \bm w_2, \ldots \bm w_N\}$ and ${\alpha}_{kj}$, $k,j=1,2\ldots,N$, henceforth known as the \emph{importance coefficients}, can be chosen in such a way that the error term in the theorem \eqref{main_bound} is minimized. Inspired by this, the following optimisation problem is proposed:
\begin{equation}
\label{eq:MtFEEL Problem}
\begin{aligned}
\min_{\bm w_1, \bm w_2, \ldots, \bm w_N  }& \min_{\bm{\alpha}}\left\{ \hat \Psi_{W, \bm \lambda, \bm \alpha} := \hat{\Phi}_{ W,\bm{\lambda},\bm{\alpha}}(\bm S) + \sum_{k=1}^{N}\gamma_{k}\parallel \bm w_k \parallel_2 +M \texttt{Pen}(\bm{\lambda}, \bm{\alpha})\right\},
\end{aligned}
\end{equation}
where the second term above  is used to regularize the neural network coefficients. The above needs to be solved in a distributed fashion ensuring that the communication overhead is low. In the following section, a federated algorithm is proposed to solve \eqref{eq:MtFEEL Problem}. 

\section{Distributed FEEL (DFL) Algorithm} \label{Algo_Section}
It is important to note that in order to solve \eqref{eq:MtFEEL Problem}, the knowledge of the discrepancy is required. However, the devices have access to data, and hence the discrepancy needs to be estimated in a distributed fashion. This estimate will be used as a proxy in \eqref{eq:MtFEEL Problem} to design the federated algorithm. In the following subsection, an algorithm to estimate the discrepancy is proposed. 
\subsection{Distributed Discrepancy Estimation (DDE)} \label{sec:dist_desc_estimation}
From the definition of the discrepancy, the DDE algorithm amounts to solving $ \sup_{\bm w \in \mathbb{R}^d} \\ \left| \mathcal{L}_k(\bm w)- \mathcal{L}_j(\bm w)\right|$ for all $k,j=1,2,\ldots, N$. Since the true average in the expression for the discrepancy is unknown, estimates of those two terms denoted by $\hat{\mathcal{L}}(\bm w, S_k)$ and $\hat{\mathcal{L}}(\bm w, S_j)$ will be used. A natural approach to solving the problem is to use a gradient ascent algorithm in a distributed manner. \textbf{Algorithm $1$} shows the distributed implementation of the gradient ascent to compute an estimate of the discrepancy. Since the discrepancy involves an absolute value, the gradient does not exist at all points. However, to circumvent this problem, a generalized (sub) gradient is used in place of the gradient (see step $8$ of the algorithm)
\begin{align} \label{eq:subgrad_ddea}
\partial_{\bm w} (\Delta_{kj}(\bm w)): = & (\nabla(\hat{\mathcal{L}}(\bm w, S_j)) - \nabla(\hat{\mathcal{L}}(\bm w, S_k))) \mathbbm{1}\{(\hat{\mathcal{L}}(\bm w, S_k) > \hat{\mathcal{L}}(\bm w, S_j))\} - \nonumber\\
& (\nabla(\hat{\mathcal{L}}(\bm w, S_k)) - \nabla(\hat{\mathcal{L}}(\bm w, S_j))\mathbbm{1}\{(\hat{\mathcal{L}}(\bm w, S_j)\leq \hat{\mathcal{L}}(\bm w, S_k))\}.
\end{align}
This follows directly from the sub-derivative of $|x|$ which is $-1$ if $x < 0$, $1$ if $x>0$ and if $x=0$, it is any point in the interval $[-1,1]$  (see \cite{boyd2003subgradient}).
As the losses are continuous random variables, the probability of the event that the losses are equal is zero, and hence the gradient in \eqref{eq:subgrad_ddea} is sufficient. 
\begin{algorithm}
\SetAlgoLined
\textproc{\small{Initialise}} discrepancies $ \hat d_{jk}$ for $ k,j \in\{1, 2, \ldots N\}$ as $1$ and $\bm w^{0}\sim\mathcal N(0,I),\; I\in \mathbb{R}^{d\times d} $\\
%\sim \mathcal{N}(\bm 0,\Sigma)$ where $\Sigma \in \mathbb{R}^{q\times q}$\\
\For{$j \in \{1, 2, \ldots, N\}$}
{
\For{$k \in \{j, \ldots, N\}$}
    {
    \For{$t \in \{1, 2, \ldots, T\}$ and $j \neq k$}
        {
            \textproc{\small{Broadcast}} $\bm w^t$  to devices  ${j}$ and ${k}$\\
            \textproc{\small{Receive}} (sub)gradients $\nabla \mathcal{\hat  L}(\bm w^t, S_j)$,  $\nabla \mathcal{\hat  L}(\bm w^t, S_k)$ and losses $\mathcal{\hat  L}(\bm w^t, S_j)$ and $\mathcal{\hat  L}(\bm w^t, S_k)$ from devices ${j}$ \text{and} ${k}$, respectively\\
            \textproc{\small{Set}} $ \hat d_{jk}= \hat d_{kj} : =|(\mathcal{\hat  L}(\bm w^t, S_j)-\mathcal{\hat  L}(\bm w^t, S_k))|$\\
            \textproc{\small{(Sub)Gradient Ascent}} using 
                $\bm w^{t+1}= \bm w^t+\eta \partial_{\bm w} (\Delta_{kj}(\bm w^t))$
        }
    }
}
\textproc{\small{Output}} all $ \hat d_{jk}$ for $ k,j \in\{1, 2, \ldots, N\}$
\caption{\textproc{\small{DDE Algorithm}} } \label{alg:dde}
\label{disc_algo}
\end{algorithm}In general, the problem is non-convex, and hence the above algorithm need not converge to the global maximum. Since this is a gradient ascent algorithm, the convergence to a local maximum follows from the standard argument \cite{boyd2003subgradient}. The time complexity of \textbf{Algorithm $1$} is polynomial of the order $\mathcal{O}(N^2dT)$. In the next subsection, using estimates of discrepancies, a distributed federated learning algorithm is developed. 

\emph{Note:} In the simulations, the number of iterations were fixed and it was observed that when the data across devices were i.i.d, $T$ was smaller than the threshold number of iterations. Although there is a scope for improvement in terms of communication complexity while estimating the discrepancy, the focus of this paper is to show that the performance of the federated algorithm can be improved using discrepancy estimate while maintaining the communication complexity to be nominal.
\subsection{Proposed DFL Algorithm}
The discrepancy estimates obtained from the DDE algorithm can be used as proxies for the true discrepancies while solving the problem in \eqref{eq:MtFEEL Problem}. One can use the classical federated algorithm to solve the problem. However, this requires exchange of gradients, which in many problems can be of very high dimension leading to a communication bottleneck. Therefore, in this section, a signed gradient method is proposed, which is different from \cite{bernstein2018signsgd}. Although a general approach of quantized gradients can used in this context, for the sake of simplicity, a simple signed gradient will be used, and relegate the analysis of quantized gradients to future work. It is important to note that the mathematical tools used here can be  extended to handle quantized gradients scenario. In the signed gradient scenario, computing the gradient of the objective in \eqref{eq:MtFEEL Problem} involves finding the sign of the gradient with respect to $\{\bm w_1, \bm w_2, \ldots \bm w_N \}$ instead of the full gradient, and the full gradient with respect to $\bm{\alpha}$. This will be a function of a relatively much lesser dimensional information. The gradient with respect to the importance coefficients $\bm{\alpha}$ does not depend on the neural network weights but depends on the discrepancy and the losses. This involves sending $\mathcal{O}(N)$ parameters while the gradient with respect to neural network weights $\bm w$ can potentially involve millions ($\gg \mathcal{O}(N)$) of parameters. The signed gradient of the objective function \eqref{eq:MtFEEL Problem} with respect to device $k$'s weights $\bm w_k$ turns out to be  
\begin{equation}
\label{eq:grad_w}
\nabla_{\bm w_k,\operatorname{sign}} \hat \Psi_{W, \bm \lambda, \bm \alpha} = \lambda_{k} \sum_{m=1}^{N} \alpha_{k m}\texttt{sign}(\hat{\bm g}_{km}) + \gamma_k \operatorname{sign}(\bm w_{k}),
\end{equation}
where the gradient at the $m^{th}$ device running model $k$ denoted by $\hat{\bm g}_{km}:= \nabla_{\bm w_k} \mathcal{\hat  L}(\bm w_k, S_m)$, and \texttt{sign}$(\hat{\bm g}_{km})$ represents the sign of the estimated gradient $\hat{\bm g}_{km}$. The full gradient of the loss $\hat \Psi_{W, \bm \lambda, \bm \alpha}$ for a device $k$, with respect to the importance coefficients $\bm \alpha_k=[\alpha_{k1} \; \alpha_{k2}\dots \; \alpha_{kN} ]$ is given by
$\nabla_{\bm \alpha_k}\hat \Psi_{W, \bm \lambda, \bm \alpha} := \left[\frac{\partial \hat \Psi_{W, \bm \lambda, \bm \alpha} }{\partial{ \alpha_{k1}}},   \; \frac{\partial \hat \Psi_{W, \bm \lambda, \bm \alpha} }{\partial{ \alpha_{k2}}}, \ldots,  \frac{\partial \hat \Psi_{W, \bm \lambda, \bm \alpha} }{\partial{ \alpha_{kN}}} \right]$, where
% $\frac{\partial \hat \Psi_{W, \bm \lambda, \bm \alpha} }{\partial{ \alpha_{km}}}= \lambda_{k}\mathcal{\hat  L}(\bm w_k,\bm S_m) + \lambda_{k} \hat{d}_{km} + \frac{M\sqrt{\frac{N}{2}\displaystyle\log\left(\frac{\mid \bm \Lambda_{\epsilon}\mid}{\delta}\right)}\frac{\left(\displaystyle\lambda_{k}^2\alpha_{k m}\right)}{n_m^2}}{\displaystyle \sqrt{\left(\displaystyle\frac{1}{2} \sum_{j=1}^{N}\displaystyle \sum_{k=1}^{N}\left(\frac{\lambda_{k}\alpha_{kj}}{n_{j}}\right)^{2}\right)}}.$
\begin{equation}
\label{eq:grad_alpha}
\frac{\partial \hat \Psi_{W, \bm \lambda, \bm \alpha} }{\partial{\alpha_{km}}}= \lambda_{k}\mathcal{\hat  L}(\bm w_k,\bm S_m) + \lambda_{k} \hat{d}_{km} + \frac{M\sqrt{\frac{N}{2}\displaystyle\log\left(\frac{\mid \bm \Lambda_{\epsilon}\mid}{\delta}\right)}\frac{\left(\displaystyle\lambda_{k}^2\alpha_{k m}\right)}{n_m^2}}{\displaystyle \sqrt{\left(\displaystyle\frac{1}{2} 
\displaystyle \sum_{k,j=1}^{N}\left(\frac{\lambda_{k}\alpha_{kj}}{n_{j}}\right)^{2}\right)}}.
\end{equation}
% In the above, $$f_{kj}: = \frac{M\sqrt{\displaystyle\log\left(\frac{\mid \Lambda_{\epsilon}\mid}{\delta}\right)}\left(\displaystyle\sum_{m=1}^{N}\lambda_{m}\alpha_{m j}\right)\displaystyle\lambda_k}{\displaystyle n_j^2\sqrt{\left(\displaystyle\frac{1}{2} \sum_{m=1}^{N}\left(\displaystyle \sum_{l=1}^{N}\frac{\lambda_{l}\alpha_{lm}}{n_{m}}\right)^{2}\right)}}.$$
% \begin{bmatrix}
% \displaystyle
%           \displaystyle\frac{1}{{n_1}^2} \\
%           \displaystyle\frac{1}{{n_2}^2} \\
%           \vdots \\
% \displaystyle           \frac{1}{{n_N}^2}
%          \end{bmatrix}
% Since the gradient with respect to $\bm{\alpha}_k$ involves simple derivative in each component, and due to lack of space, an expression for the same is omitted.

%\begin{figure}[h]
%    \centering
%    \includegraphics[scale=0.17]{Images/Overview001.png}
%\vspace{-0.6cm}
%\caption{The figure depicts the general update process of the DFL algorithm over a wireless channel}  %\label{fig:fed_model1}
%\end{figure}
In particular, each device $m$ sends the signed gradients $\texttt{sign}(\hat{\bm g}_{km})$ (from equation \eqref{eq:grad_w}) and losses $\mathcal {\hat L}(w_m, S_k)\; \forall\; k = 1, 2,\ldots, N $ to the BS. The BS aggregates these according to \eqref{eq:grad_w} and \eqref{eq:grad_alpha} to get the estimated gradients $\nabla_{\bm w_k,\operatorname{sign}} \hat \Psi_{W, \bm \lambda, \bm \alpha}$ and $\nabla_{\bm \alpha_k} \hat \Psi_{W, \bm \lambda, \bm \alpha}$. These estimates are used in obtaining an improved estimate of the optimal weights $\bm w_k$ and importance coefficients $\alpha_{kj}$ for all $k,j = 1, 2, \ldots, N$. In order to satisfy the constraint that $\sum_{j=1}^N \alpha_{kj} = 1$, the importance coefficient needs to be projected onto the simplex $\Delta_N$ (see step $8$ of the \textbf{Algorithm $2$}). Note that the gradient of \eqref{eq:MtFEEL Problem} with respect to $\bm{\alpha}$ depends on the estimated discrepancy. Computing this gradient is straightforward, and hence not explicitly mentioned in the algorithm. The convergence analysis of \textbf{Algorithm $2$} is discussed in the next section. 
\begin{algorithm}
\SetAlgoLined
%\algsmall{Get} discrepancy $\bm  d_{kj} \forall k,j \in\{1\dots N\}$ from \algsmall{Discrepancy}\\
\algsmall{Initialise} $\bm \alpha_k^0 \in \Delta_N$, $\bm w_k \in \R^d$ for $ k = 1,2,\ldots, N$ \\
\For{$t = 1, 2, \ldots, T$}
{
    \algsmall{Broadcast} $\bm w_j^t $ to the device $j = 1,2,\ldots, N $\\
    \For{ devices $k = 1, 2, \ldots, N$}
    {   
        \algsmall{Get} $\texttt{sign}(\hat{\bm g}_{mk})\; \text{ for } m= 1,2,\ldots, N$\\
        \algsmall{Get} $\hat{\mathcal{L}}(\bm  w_m,S_k)\; \text{ for } m = 1,2,\ldots, N$ \\
    %    \algsmall{Get} from \cl{k} all losses \\
    }
    \algsmall{Gradient Descent} step on $\bm w_k \; \text{ for } k = 1,2,\ldots, N$:\newline $\bm w_k^{t+1}= \bm w_k^t-\eta \nabla_{\bm w_k,\operatorname{sign}} \hat \Psi_{W^{t}, \bm \lambda, \bm \alpha^t} $\\
    \algsmall{Gradient Descent with Projection}  step on $\bm \alpha_k \; \text{for} \; k = 1,2, \ldots, N$:\newline 
    $\bm \alpha=\bm \alpha_k^t - \mu \nabla_{\bm \alpha_k} \hat \Psi_{W^{t+1}, \bm \lambda, \bm \alpha^t}$\newline
    $\bm \alpha_k^{t+1}= \underset{{\bm x \in \Delta_N}}{\operatorname{argmin}}\parallel \bm x-\bm \alpha\parallel_1$
}
\caption{Proposed DFL (MtFEEL): Input discrepancy from \textbf{Algorithm $1$}}
\label{MtFEEL_ALGO}
\end{algorithm}

\section{Convergence Analysis} 
\label{sec:convergence_analysis}
In this section, the convergence analysis of  \textbf{Algorithm $2$} is presented. In order to prove one of the main results on the convergence, the following standard assumptions are made. 
 
\begin{assumption} \label{assumption1}
(Boundedness {\cite{hazan2017efficient}}): The loss function $l(h_{\bm w}(\bm x),y)$ is assumed to be bounded i.e., $l(h_{\bm w}(\bm x),y)\leq B< \infty,$ for all $\bm w \in \mathbb{R}^d$ and any $\bm x \in \mathcal{X}$ and $y \in \mathcal{Y}$. 
\end{assumption}
Assumption 1 implies that $\Phi_{W,\bm \lambda, \bm \alpha}(\bm S)\leq B$, which follows from the facts that $\SUM{N}{k=1}\lambda_k=1$ and $\SUM{N}{j=1}\alpha_{kj} =1,$ $k=1,2, \ldots, N$. Next, the gradient is assumed to be smooth (see \cite{bernstein2018signsgd}).
%This assumption guarantees convergence to a stationary point.
\begin{assumption}
\label{assumption 2}
($\beta-$ Smoothness): The function $l(h_{\bm w}(\bm x),y)$ is assumed to be $\beta-$smooth in $\bm w$, i.e.,  
$\mid \left(\nabla l(h_{
\bm w_{1}}(\bm x),y)_i-\nabla l(h_{\bm w_{2}}(\bm x),y)_i\right)\mid  \leq L\mid (\bm w_1- \bm w_2)_i \mid$ for any $\bm w_1, \bm w_2 \in \mathbb{R}^d, \bm x \in \mathcal{X}, y \in \mathcal{Y}, i= 1,2, \ldots, d,$ and $0<L_i<\infty $ for all $i$.
\end{assumption}
 Note that the above assumption implies that
\[
\left|l(h_{\bm w_1}(\bm x),y)-\left[l(h_{\bm w_2}(\bm x),y)+{\nabla l(h_{\bm w_{1}}(\bm x),y)}^{T}(\bm w_1- \bm w_2)\right]\right| \leq \frac{1}{2} \sum_{i} L_{i}\left(w_{1i}- w_{2i}\right)^{2},
\] 
for any $\bm w_1, \bm w_2 \in \mathbb{R}^d, \bm x \in \mathcal{X}$ and $ y \in \mathcal{Y}.$
\begin{assumption}
\label{assumption3}
(Bounded Variance):  Assume that every component of the estimated gradient, i.e., $\hat g_{kj,i}$ is an unbiased estimate of the true gradient $g_{kj,i}$, i.e., $\mathbb{E}\left[\hat g_{kj,i}\right] = g_{kj,i}$, \; for $k, j = 1,2,\ldots, N$ and $i = 1,2,\ldots, n_j$. Further, the variance of every component is bounded by some $\sigma_{kj,i}^2< \infty $, i.e.,
$\mathbb{E}\left[{(\hat g_{kj,i} - g_{kj,i})}^2\right] \leq \sigma_{kj,i}^2.$
\end{assumption}

Since $\bm{\alpha}_k$ should satisfy the constraint that $\sum_{i=1}^N \alpha_{ki} = 1$, the gradient descent step is followed by a projection. The following definition comes in handy while deriving the convergence results for the proposed algorithm. 

\begin{definition}
(\textbf{Projected gradient}  \cite{hazan2017efficient})  Let $\Psi: \mathcal K \rightarrow \mathbb{R}$ be a function on a closed convex set $\mathcal K \subseteq \Delta_{N}$. The projected gradient of $\bm z \in \mathbb{R}^d$ with respect to $\Psi$ denoted $\nabla_{\mathcal K, \bm z} \Psi : \mathcal K \rightarrow \mathbb{R}^N $ is defined as
\begin{equation}
\nabla_{\mathcal K, \bm z} \Psi : = \frac{1}{\mu}(\bm z - \Pi_{\mathcal K}[\bm z - \mu\nabla_{\bm z} \Psi(\bm z)]),
\end{equation}
where $\Pi_{\mathcal K}(\bm z)=\underset{{\bm x \in \Delta_N}}{\operatorname{argmin}}\parallel \bm x- \bm z\parallel$ is the projection operator, and any $\mu > 0$.
\end{definition}
Note that if $\Pi_{\mathcal K}(\bm z)= \bm z $, then the above coincides with the gradient. Further, gradient update using the above ensures that 
$\sum_{j=1}^{N}\alpha_{kj} = 1$. In order to prove convergence of the \textbf{Algorithm $2$}, it suffices to prove that $\bm w_k$ and $\bm \alpha_k $ converges for all $k=1,2, \ldots, N$. Therefore $\Psi_{W,\bm \lambda, \bm \alpha}$ for the $k$-th component is written as,
\begin{equation}
\Psi_{\bm w_k, \lambda_k, \bm \alpha_k} = \Phi_{\bm w_k,{\lambda_k},\bm{\alpha_k}}(\bm S) + \texttt{Reg}({\lambda_k}, \bm{\alpha_k}) + \lambda_{k} \sum_{j=1}^{N}\alpha_{kj}{ d}_{kj}, 
\label{MtFEEL_k}
\end{equation}
where $\Phi_{\bm w_k,{\lambda_k},\bm{\alpha_k}}(\bm S): = \lambda_k \sum_{j=1}^{N} \alpha_{kj} \mathcal{L}_j(\bm w_k),$ and 
$$\texttt{Reg}({\lambda_k}, \bm{\alpha_k}): = \frac{M}{N}\sqrt{\frac{N}{2} \sum_{j=1}^{N}\sum_{k=1}^{N}\left(\frac{\lambda_{k}\alpha_{k j}}{n_{j}}\right)^{2}\log\left(\frac{\mid\bm \Lambda_{\epsilon}\mid}{\delta}\right)}.$$ It is important to note that the term corresponding to the $L_2$-regularizer $\gamma_k \left \vert \bm{w}_k \right \vert$ is ignored, i.e., $\gamma_k=0$ in order to prove the convergence result. However, the proof can be easily extended to the case of $\gamma_k \neq 0$. The proof of convergence requires the objective $ \Psi_{\bm w_k, \lambda_k, \bm \alpha_k}$ in (\ref{MtFEEL_k}) to be Lipschitz function of $\bm \alpha_k$ and $\bm w_k$. This is the essence of the following Lemmas. 
% The proof of convergence requires the objective $\hat \Psi_{W, \bm \lambda, \bm \alpha}$ in (\ref{eq:MtFEEL Problem}) to be Lipschitz function of $\bm \alpha$. Since $\hat \Psi_{W, \bm \lambda, \bm \alpha}$ is linear in $\bm \alpha$, it suffices to prove that $\texttt{Pen}(\bm{\lambda}, \bm{\alpha})$ is Lipschitz in $\bm \alpha$. The term  $\texttt{Pen}(\bm{\lambda}, \bm{\alpha})$ has two terms, one is linear in $\bm \alpha$, and the other term $\texttt{Reg}(\bm{\lambda}, \bm{\alpha})$ is non-linear in $\bm \alpha$. Hence it suffices to prove the following Lemma.
% In order to prove convergence of the algorithm, it suffices to prove that $\bm w_k$ converges for all $k=1,2, \ldots, N$. Equivalently, it is enough to show that the gradient of
% \begin{equation}
% \Psi_{\bm w_k, \lambda_k, \bm \alpha_k} = \Phi_{\bm w_k,{\lambda_k},\bm{\alpha_k}}(\bm S) + \gamma_{k}\parallel \bm w_k \parallel_2 + \texttt{Reg}({\lambda_k}, \bm{\alpha_k}) + \lambda_{k} \sum_{j=1}^{N}\alpha_{kj}{ d}_{kj}, 
% \end{equation}
% which corresponds to the $k$-th component of $\Psi_{W,\bm \lambda, \bm \alpha}$, converges. Further,  \\
% $\Phi_{\bm w_k,{\lambda_k},\bm{\alpha_k}}(\bm S): = \lambda_k \sum_{j=1}^{N} \alpha_{kj} \mathcal{L}_j(\bm w_k),$ and 
% $\texttt{Reg}({\lambda_k}, \bm{\alpha_k}): = \frac{\texttt{Reg}(\bm{\lambda}, \bm{\alpha})}{N}$.

\begin{prop} 
\label{lem:lips_reg}
The function $\texttt{Reg}(\lambda_k, \bm{\alpha_k}): = \frac{M}{N}\sqrt{\frac{N}{2} \sum_{j=1}^{N}\sum_{k=1}^{N}\left(\frac{\lambda_{k}\alpha_{k j}}{n_{j}}\right)^{2}\log\left(\frac{\mid\bm \Lambda_{\epsilon}\mid}{\delta}\right)}$ is Lipschitz in $\bm \alpha_k$ with Lipschitz constant $\beta^{\prime} : = \frac{M}{\sqrt{2N}}\sqrt{\log\left(\frac{\mid \bm \Lambda_{\epsilon}\mid}{\delta}\right)}$.
\end{prop}
\emph{Proof:} The proof is provided in Appendix \ref{Validating}. \qed 

\begin{prop} \label{lem:lips_psihat}
The function $ \Psi_{\bm w_k,\lambda_k, \bm \alpha_k}$ is Lipschitz in $\bm \alpha_k$ with Lipschitz constant $\beta: = \beta^{\prime} + 2 \lambda_kM $ and Lipschitz in $\bm w_k$ with Lipschitz constant $\lambda_k d$.
\label{lipconstants}
\end{prop}

\emph{Proof:} The proof is provided in Appendix \ref{Prop2}. \qed 

The following theorem uses the above results and definitions to show that \textbf{Algorithm $2$} converges.

\begin{theorem}(Convergence of \textbf{Algorithm $2$}).
\label{Theorem 2}
After $T$ iterations, choosing the learning rates $\eta^t = \frac{1}{\sqrt{T}}, \; \mu^{t} = \frac{1}{\sqrt{T}}$ and the batch size $n_t = T$, the following holds:
\begin{eqnarray} \label{eq:conv_alg2}
% \mathbb{E}\Bigg[\frac{1}{T}\sum_{t=0}^{T-1}\left(\lambda_{k}^{2}\sum_{m=1}^{N}  \alpha_{k m}^{t+1}\left\|\bm g_{k m}^t\right\|_{1}\right)\Bigg] + \left(1-\frac{\beta }{2\sqrt{T}}\right) \mathbb{E}\left[\frac{1}{T}\sum_{t=0}^{T-1}\left\|\nabla_{\mathcal K, \bm \alpha_{k}} \Psi_{\bm w_k^t, \lambda_k, \bm \alpha_k^t} \right\|^{2}\right] \leq \nonumber\\
\mathbb{E}\left[\frac{1}{T}\SUM{T-1}{t=0}\Delta_k^t\right] \leq
\frac{1}{\sqrt{T}}\Bigg(2\lambda_{k}^{2} \sum_{m=1}^{N}\parallel \bm \sigma_{k m}\parallel_{1} +  \frac{\parallel L \parallel_{1} \lambda_{k}^{2}}{2} + \Psi_{\bm w_k^0, \lambda_k, \bm \alpha_k^0} - \Psi_{k}^{*}\Bigg),
\label{eq: theorem2}
\end{eqnarray}
where $\Delta_k^t:=\left(\lambda_{k}^{2}\sum_{m=1}^{N} \alpha_{k m}^{t+1}\left\|\bm g_{k m}^t\right\|_{1}\right) + \left(1-\frac{\beta }{2\sqrt{T}}\right)\left[\left\|\nabla_{\mathcal K, \bm \alpha_{k}} \Psi_{\bm w_{k}^{t}, \lambda_k, \bm \alpha_{k}^{t}}\right\|_2^{2}\right]$.
\label{theorem 2}
\end{theorem}
\emph{Proof:} The proof is provided in Appendix \ref{thrm2}.\\
Note that as $T \rightarrow \infty$, the right hand side goes to zero. 
In other words, each term on the right hand side of (\ref{eq: theorem2}) can be made arbitrarily small by choosing appropriately large $T$. 
Since the first term corresponds to the average gradient of $\Psi_{\bm w_k, \lambda_k, \bm \alpha_k } $ with respect to $\bm w_k$ scaled by $\lambda_k$, and the second term corresponds to the gradient of $\Psi_{\bm w_k, \lambda_k, \bm \alpha_k}$ with respect to $\bm \alpha_k$, there exists a time $t$ beyond which the gradient of $\Psi_{\bm w_k, \lambda_k, \bm \alpha_k } $ is small. 
This shows that rate of convergence is $\mathcal{O}\left(\frac{1}{\sqrt{T}}\right)$ similar to \cite{bernstein2018signsgd}. 
In the next section, using the results of this section, a convergence result of the proposed algorithm when the channel between the edge device and the BS is a noisy wireless channel is presented.

\vspace{-5mm}
\section{Convergence analysis in a Wireless Channel}
\label{sec:rayleigh}
The above analysis holds good in the presence of an error free channel. For a more pragmatic approach,  the  following  analysis  on  a  noisy  channel  is  presented. The channels between each device $k$ and the BS is considered to be a single tap Rayleigh flat fading channel, with channel coefficient $h_k$ as defined in section \ref{subsec:channel_model}.\footnote{The time slot index $t$ in $h_k(t)$ is ignored, and is understood from the context.} It is assumed that the entries of $\operatorname{sign}(\hat g_{ki})$ is i.i.d with the probability of $1$ being $q$. It is clear that if $d$ bits are sent without any errors, then the convergence is guaranteed, as in Theorem \ref{thm: Theorem3}. However, in the wireless channel, it is expected to have errors, which depends on the SNR of the channel. Therefore, it is important to investigate the impact of the SNR on the convergence. Towards this, define the outage event as $\mathcal{O}:= \left\{d \leq \log_2\left (1+\displaystyle \frac{|h_k|^2 {P}_k}{B\sigma^2}\right)\right\}$, where the maximum transmissible power by a user is ${P}_k$ and the channel noise variance is $\sigma^2$ for channel bandwidth $B$.\footnote{The symbol $\mathcal{O}$ is used to represent both outage as well as ``order of". It should be clear depending on the context.} It is assumed that in the case of outage event, the errors are bound to happen, and hence the communication is said to have failed. Strictly speaking, even in the case of outage, it is possible that a few bits will get through without any errors, which can be used to move roughly in the direction of gradient. This can potentially help in improving the convergence. However, for the sake of simplicity, the above case is ignored. Further, the discrepancy estimates assume error free channel. As mentioned earlier, the impact of scheduling on the convergence is ignored. In this setting, the following theorem characterizes the convergence of the proposed algorithm under fading channel. %Defining 
\begin{theorem}
\label{thm: Theorem3}
In a Rayleigh fading up-link channel with $SNR_k:= \frac{P_k}{B \sigma^2}$ and bandwidth $B$ at the device $k$, by choosing the learning rates, and batch size as in Theorem \ref{theorem 2}, the following bound holds good
\[
\frac{1}{T} \sum_{t=1}^T \mathbb{E}[\Delta_k^t] \leq \frac{U (\displaystyle 2^{\frac{d}{B}}-1)}{SNR_k}+
 \frac{1}{\sqrt{T}}\Bigg(2\lambda_{k}^{2} \sum_{m=1}^{N}\parallel \bm \sigma_{k m}\parallel_{1} +  \frac{\parallel L \parallel_{1} \lambda_{k}^{2}}{2} + \Psi_{\bm w_k^{0}, \lambda_k, \bm \alpha_k^{0}} - \Psi_{k}^{*}\Bigg),
\]
%  if the SNR corresponding to the device $k$ satisfies
% \begin{equation*}
%     \begin{aligned}
%     \displaystyle
%     \text{SNR}_{k} \geq \frac{\left( 2^{d/B}-1\right)}{  \ln{\sqrt{T}} - \ln{\sqrt{T-1}}}
%     \end{aligned}
% \end{equation*}
where $U: = \beta^{\prime} + 2 \lambda_kM + \lambda_kd$.
\end{theorem}
It can be observed from the above theorem that the average is small provided that the quantity $\mathcal{O}\left(\max\left\{\frac{1}{SNR_k}, \frac{1}{\sqrt{T}}\right\}\right)$ is small. This ensures that there exist a $t$ for which $\Delta_k^t$ is small, making the sum of the gradients of $\Psi_{\bm{w}_k,\bm{\alpha}_k,\lambda_k}$ with respect to $\bm{\alpha}_k$ and the neural network weights $\bm{w}_k$ small, ensuring that the solution is close to a sub-optimal minimum. However, for a fixed $SNR_k$, it is useless to train for more that $\mathcal{O}(SNR_k^2)$ number of iterations. Thus, the SNR of the transmission acts as a bottleneck while training in the fading channel scenario. In addition, the higher the Bandwidth, lower the number of iterations required. These observations are made in the experimental results as well, which is detailed in the next section. 
% \vspace{-5mm}
\section{Experimental Results} \label{sec:expresults}
The experimental setup uses the MNIST handwriting data set 
% \cite{lecun-mnisthandwrittendigit-2010} 
to emulate a federated learning scenario with $N=30$ devices. More specifically, 
% to emulate an FL scenario with devices having data of ``similar" and heterogeneous distributions,
three cohorts of devices denoted $A$, $B$ and $C$ were constructed. Further,  $12$ devices were assigned to each cohorts $A$ and $B$ while $6$ devices were allocated to the cohort $C$. All the devices were equipped with a neural networks capable of performing a $10$ class prediction on the images. As a part of training, all the devices in cohort $A$ had the digits $0-5$, devices in $B$ were given digits $6-9$ and devices in $C$ were given digits $3-7$. This emulates the heterogeneous data set across cohorts while having homogeneous data within each cohort. It is important to note that the cohorts are \emph{unknown}. There were around $100$ samples in every device out of which $20$ were used for training and $80$ for testing. This mimics edge devices with very less data and the larger number of samples for testing was set to gain insights on the generalisation capabilities of the algorithm.
In order to compare the proposed algorithm with some baseline performance, each device was also trained using just local data. The proposed algorithm is also compared with the popular FedSGD and FedAVG (see \cite{fedsgdalgo} and \cite{mcmahan2017communication}) algorithms. In the following section, the performance of the proposed algorithm when the channel is noise free is presented. 
\vspace{-5mm}
\subsection{Perfect Uplink-Channel}
\label{subsec:perfect_channel}
\begin{figure}[!ht]
\begin{minipage}[b]{0.5\linewidth}
  \centering
  \centerline{\includegraphics[scale=0.12]{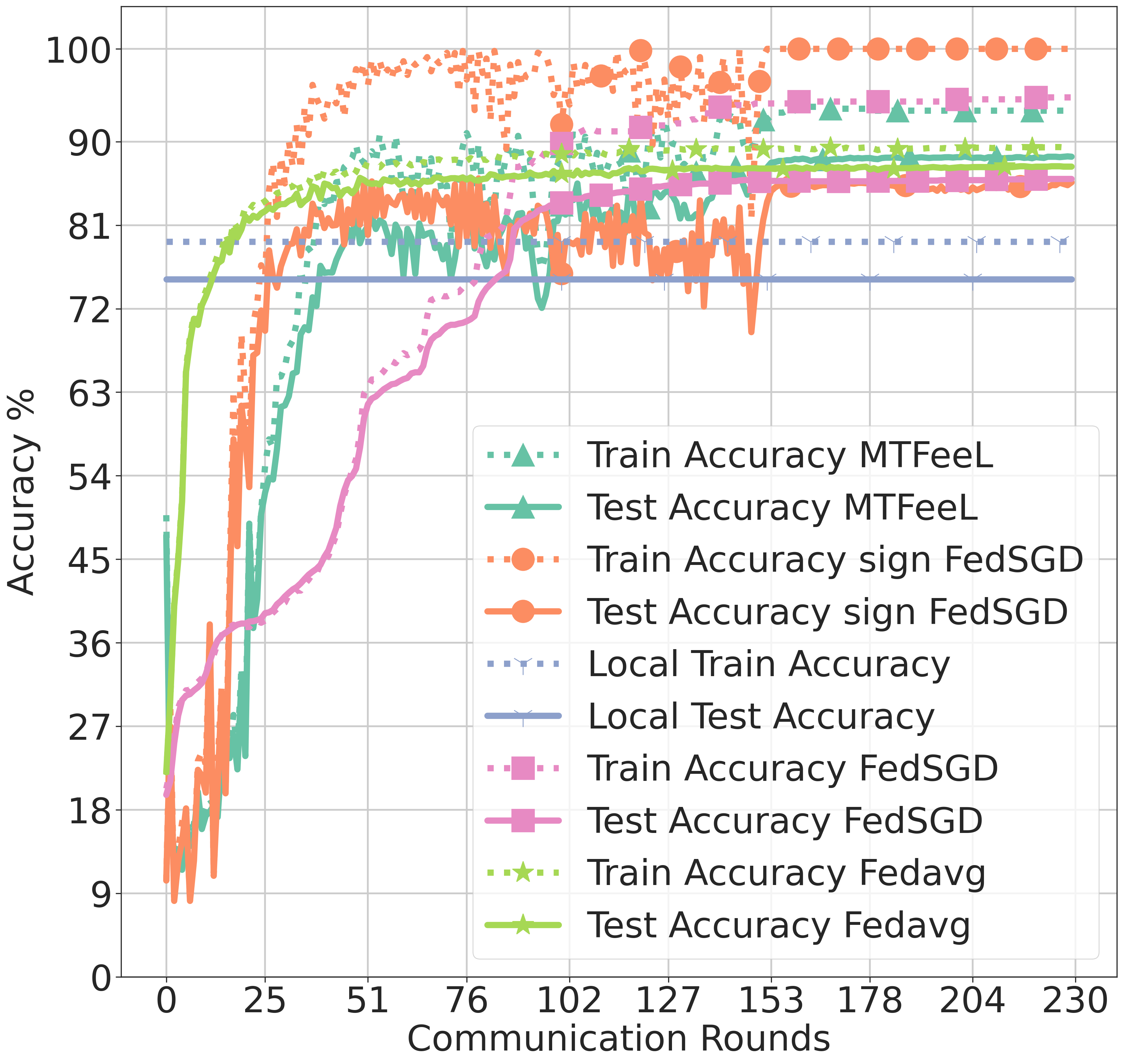}}
\centerline{(a) Performance from start to finish.}
\end{minipage}
\hfill
\begin{minipage}[b]{0.5\linewidth}
  \centering
\centerline{\includegraphics[scale=0.12]{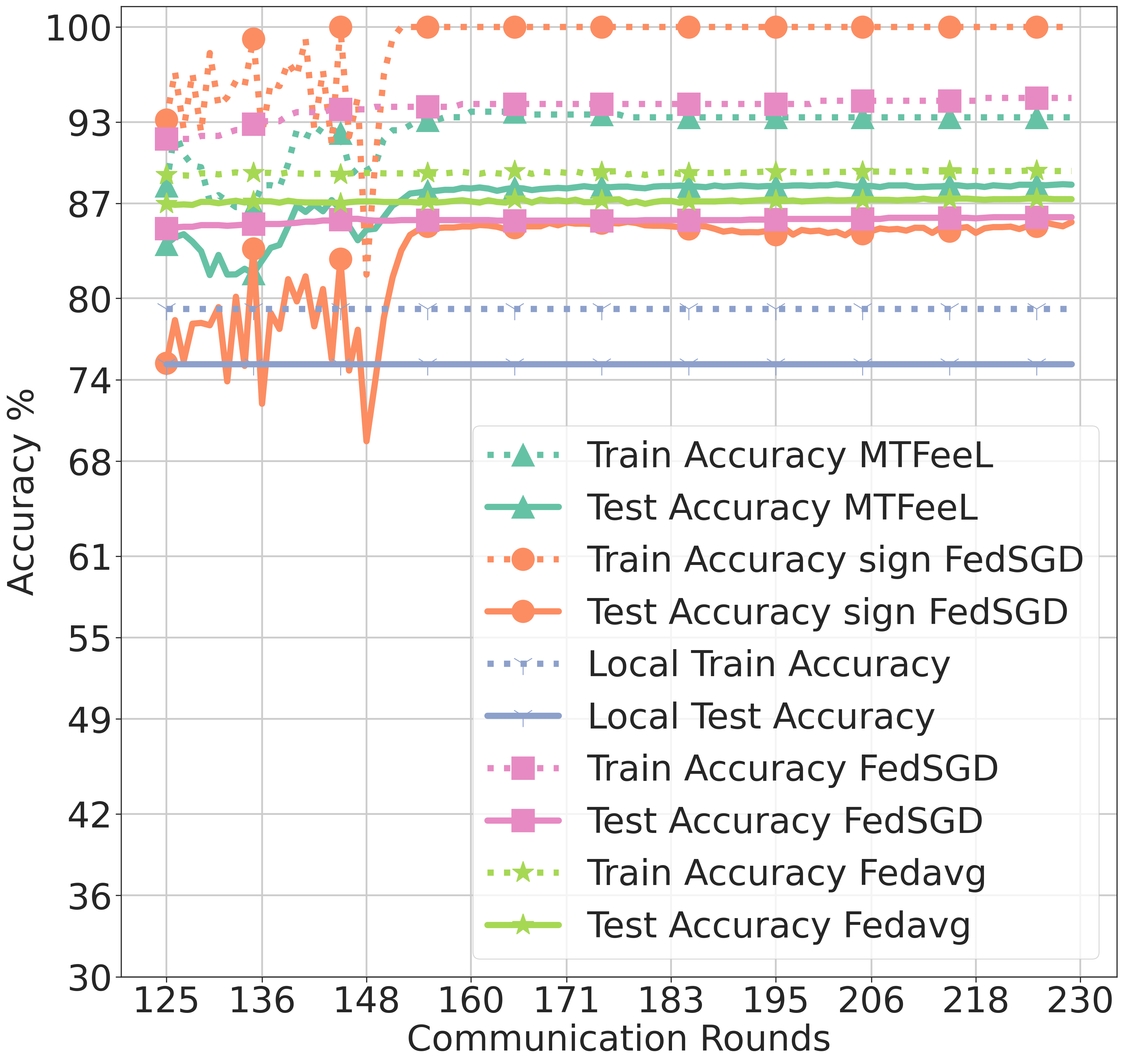}}
%  \vspace{1.5cm}
  \centerline{(b) Performance towards the end.}\medskip %\label{fig:exp_errorfree_lastcommround}
\end{minipage}
\caption{Accuracy in the case of Error free channel.}
\label{fig:exp_res}
\end{figure}

\begin{figure}[!hb]
\begin{minipage}[b]{0.35\linewidth}
  \centering
  \centerline{\includegraphics[scale=0.115]{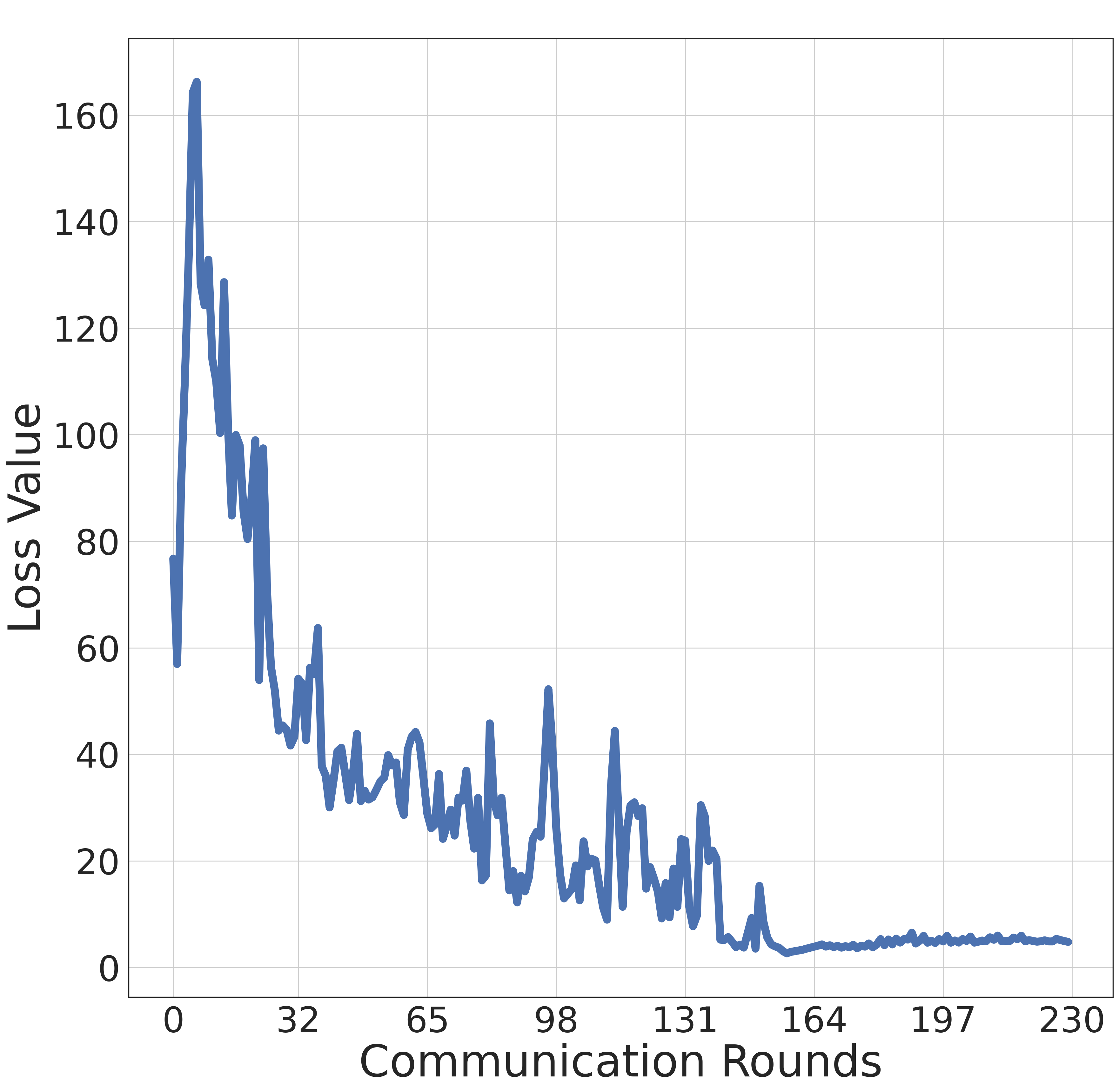}}
\centerline{(a) MtFEEL loss}\medskip
\end{minipage}
\begin{minipage}[b]{0.3\linewidth}
  \centering
  \centerline{\includegraphics[scale=0.115]{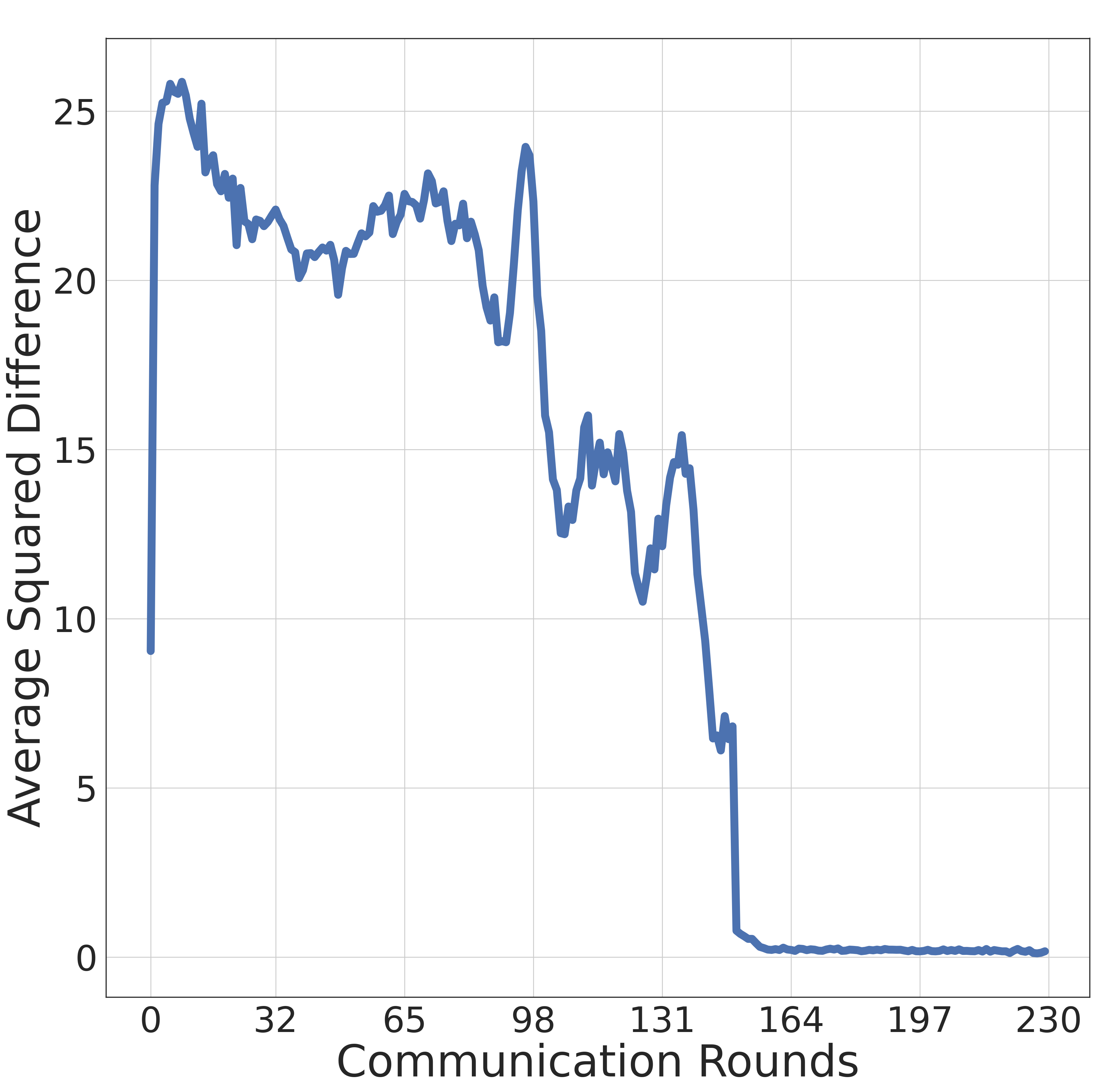}}
\centerline{(b) Neural Network Weights}\medskip
\end{minipage}
\begin{minipage}[b]{0.3\linewidth}
  \centering
  \centerline{\includegraphics[scale=0.115]{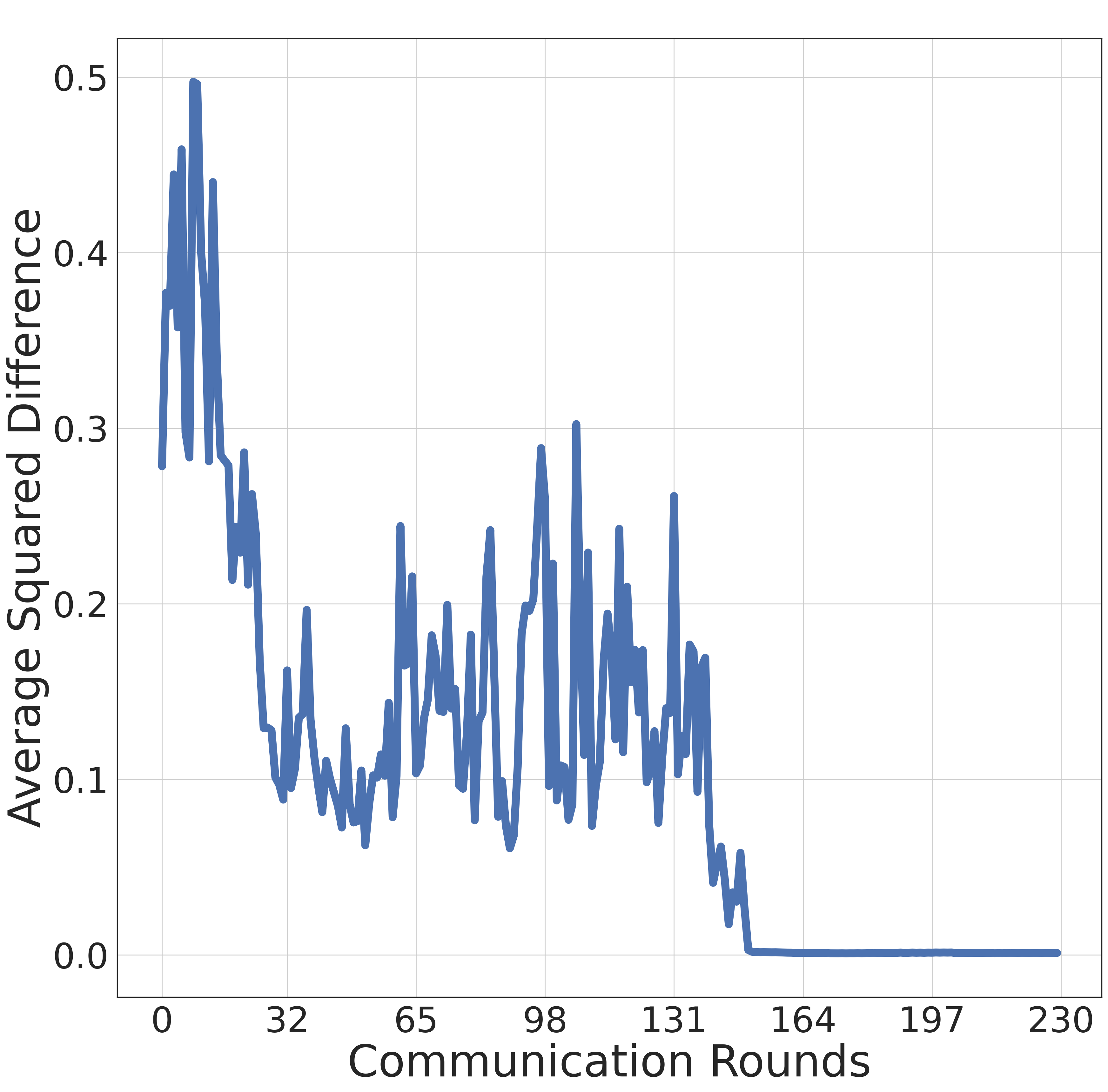}}
\centerline{(c) Importance Coefficients }\medskip
\end{minipage}
\caption{MtFEEL parameters across communication rounds in an error free channel.}
\label{fig:exp_res_parameters}
\end{figure}

In this scenario, the communication between devices and the BS occurs in an error free regime.
To begin with, the discrepancy is computed for each pair of devices using \textbf{Algorithm \ref{alg:dde}} described in Sec.~\ref{sec:dist_desc_estimation}. Using this in \textbf{Algorithm \ref{MtFEEL_ALGO}}, the respective neural network weights are computed. Figures \ref{fig:exp_res}(a) and \ref{fig:exp_res}(b) present the accuracy of various algorithms during training and testing phases versus communication rounds. Here, in each communication round, the gradient is updated using step $8$ of the \textbf{Algorithm} \ref{MtFEEL_ALGO}. It is observed that both FedSGD and sign FedSGD perform similarly albeit the latter being more unstable in the initial few rounds of communication. MtFEEL also acts unstably, but to a much lesser degree as depicted in figure \ref{fig:exp_res}(a). This instability can be attributed to the fact that the quantization (binarization to be precise) errors in the initial stages will be large as the magnitude of the gradients are much larger than $1$. The fluctuations eventually cease after around $150$ communication rounds as the gradient approaches the local minimum. It is important to note that the testing accuracy of the proposed algorithm is better than local training, FedAvg and FedSGD. On the other hand, the training accuracy of the proposed algorithm is inferior compared to FedAvg and FedSGD; this is attributed to the over-fitting of the algorithms. The superior performance of the proposed algorithm is due to the fact that the discrepancy for all the devices in a cluster is small. Hence the solution to the optimization problem results in close to equal  weights being allocated to the devices within a cluster and approximately zero weights across clusters during the learning phase. In summary, it was observed that for devices in a cluster, the algorithm converged to FedSGD, as expected. Recall that the cluster devices are unknown, and the algorithm managed to learn them quite well. 
% The algorithm takes around 30 communication rounds to outperform local training, unlike FedSGD which needs nearly 90 rounds. This can be attributed to the fact that gradient components with magnitude less than $1$ would need a smaller learning rate MtFEEL favours only those devices with lower discrepancies, thereby preventing the influence of less similar nodes from dictating the optimum.
\\
The MtFEEL average loss (see \eqref{eq:MtFEEL Problem}) versus communication rounds is depicted in Fig.~\ref{fig:exp_res_parameters} (a). It reaches its minimum at around $150$ iterations, which is when no more significant gradient descent steps are being taken. This is also shown in Figs.~\ref{fig:exp_res_parameters} (b)  and (c) which depict $\parallel\bm w_k^{t+1}-\bm w_k^{t}\parallel_2^2$  and $\parallel\bm \alpha_k^{t+1}-\bm \alpha_k^{t}\parallel_2^2$ averaged across all devices $k$, varying across communication rounds. This confirms the convergence of the MtFEEL algorithm. These experiments demonstrate a proof of concept and more elaborate experiments using different data sets is relegated to future work. In the next subsection, the performance of the proposed algorithm under fading channel is presented. 
\subsection{Noisy Uplink Channel}
\begin{figure}[!ht]
\begin{minipage}[b]{0.5\linewidth}
  \centering
  \centerline{\includegraphics[scale=0.12]{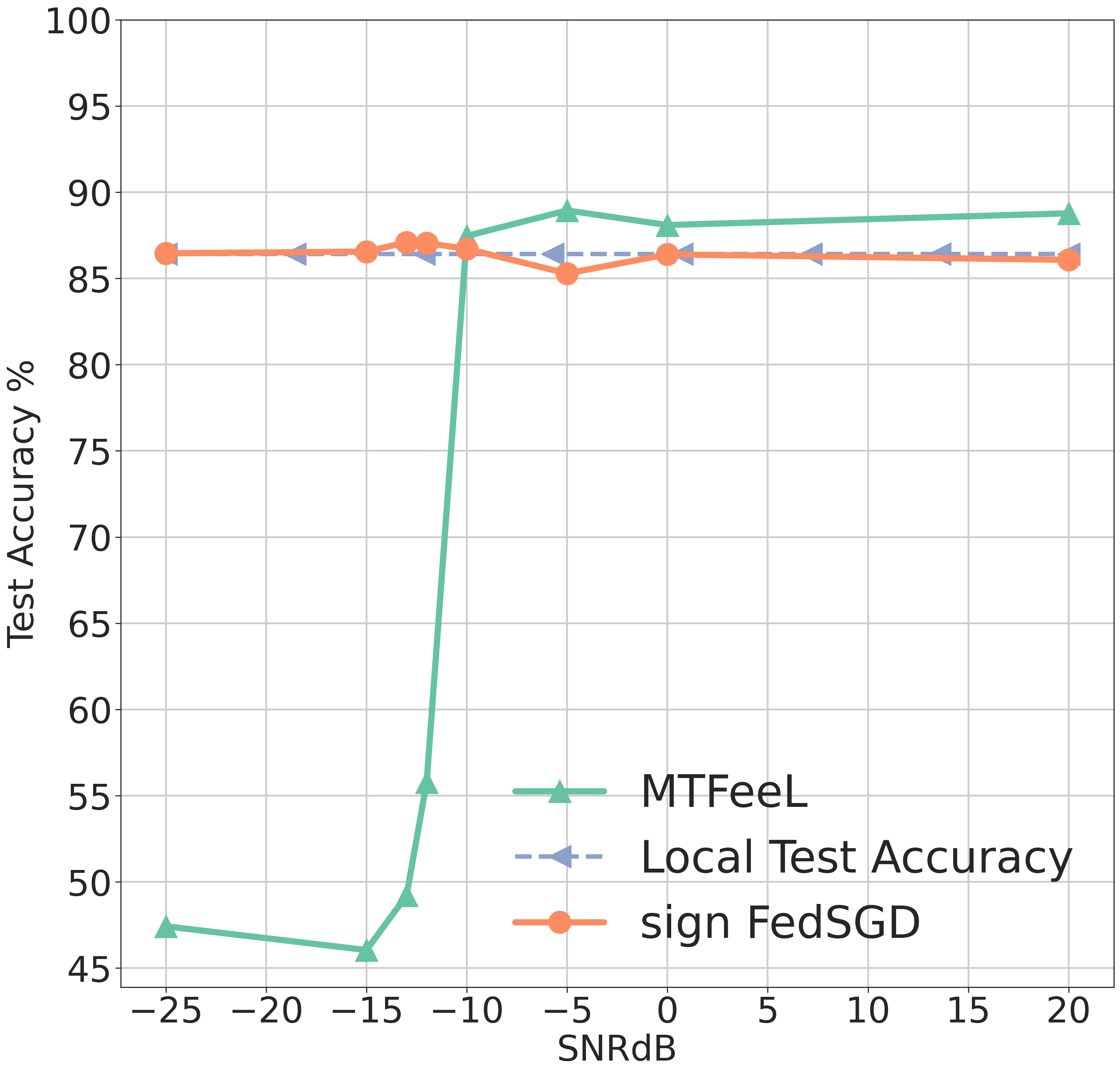}}
\centerline{(a) Rayeigh Fading Channel}\medskip
\end{minipage}
\begin{minipage}[b]{0.5\linewidth}
  \centering
  \centerline{\includegraphics[scale=0.12]{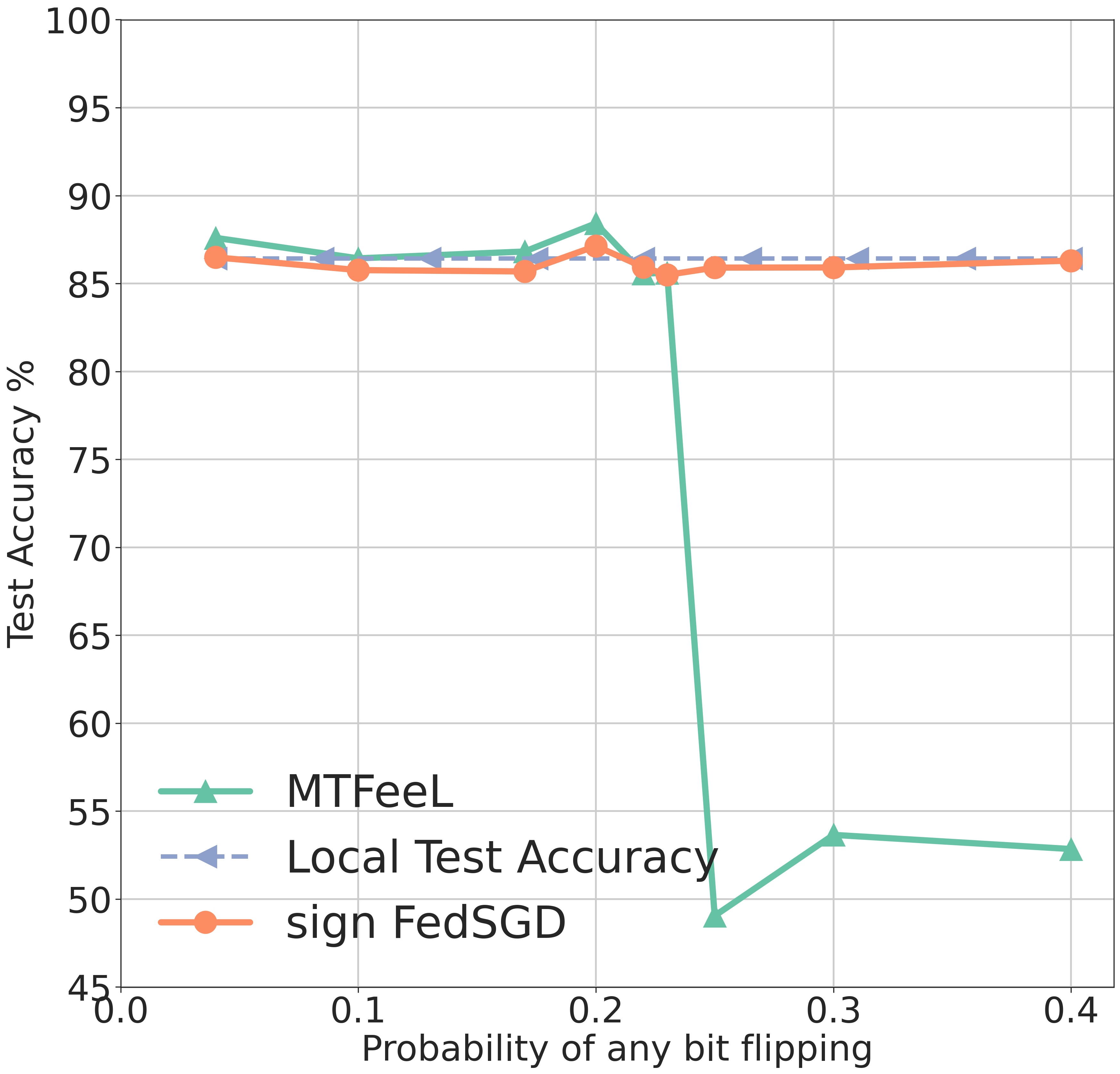}}
\centerline{(b) Bit flipping channel}\medskip
\end{minipage}
\caption{Channel with Error.}
\label{fig:Error_chann}
\end{figure}

The MtFEEL algorithm is sensitive to the loss values incurred by a model when run on different devices. 
% When the initial gradients steps are erroneous, this sensitivity to the loss values leads to unreliable estimates of the importance coefficients $\bm \alpha$, which in turn can result in poor performances.
The estimates of the importance coefficients  $\bm \alpha$ being sensitive to these loss values, can be unreliable if the initial gradients are erroneous.
It is observed that for a given uplink $SNR_k$ for device $k$, increasing the number of communication rounds may not help in improving the performance. In fact, beyond some threshold on $T$ for a given $SNR$, the performance can degrade due to the fact that the gradient is bounded away from zero as $T$ increases, as indicated in Theorem \ref{thm: Theorem3}. This is empirically observed in Fig.~\ref{fig:Error_chann} (a). For $SNR_k < -10$dB, the performance degrades, and results in very poor accuracy. In addition to the above, a bit flipping model is also considered, where each component of the gradient is independently flipped to $1$ (or $-1$) with a probability of $p$ (or $1-p$). The accuracy versus $p$ is plotted in Fig.~\ref{fig:Error_chann} (b). It is clear that beyond a threshold on $p$ ($p > 0.2$), the performance of the proposed algorithm improves, an observation in line with fading channel case. In summary, it is better to use the proposed algorithm in most practical regimes of interest, while the classical FedSGD approaches are better for very low SNR due to its robustness for large number of errors. %From a practical standpoint, the rare occasions of channel bit flip probabilities $0.2<$ would benefit from the signed FedSGD approach. The results of such an analysis is summarised in figure \ref{fig:Error_chann}.
The reason behind this is that the signed FedSGD doesn't attempt to find similar devices to aggregate, and naively considers all devices homogeneous, and can maintain its robustness for relatively higher error probabilities.
% Bitflip of 0.2 is high. From practical standpioint, it is better to use FedSGD
\vspace{-4mm}
\section{Conclusions}
\label{sec:conclusion}
This work proposed a distributed FL algorithm across multiple devices that results in custom neural network for each device. In particular, every device learns its model with the help of other ``similar" devices by sending signed gradient information to a central BS.
Each device aims to minimise an estimate of {a proposed loss} using a weighted average of the empirical losses across devices. These weights, called the importance coefficients, are dependent on the similarity of data distributions between any pair of clients in the network.
This loss function is minimised by computing neural network weights tailor made for each device. Theoretical guarantees on the proposed estimation method are provided, and an algorithm is devised to compute the importance coefficients and the neural network weights across devices.
The guarantee is depends on the weighted average of the losses, a notion called discrepancy which is a measure of the dependency of the data across devices with respect to the loss function, and a penalty term.
% The discrepancy is a measure of the dependency of the data across devices with respect to the loss function. 
An algorithm is proposed to estimate this discrepancy in a distributed fashion.
The FL algorithm was shown to converge at the rate of $1/\sqrt{T}$, where $T$ is the number of communication rounds when no errors are present in the communication links in the network. In the case of a Rayleigh flat fading channel, the convergence of the algorithm is shown to depend on the SNR and $1/\sqrt{T}$.
In particular, it was shown that the convergence is limited by the SNR, i.e., at low SNR, the increase in communication rounds $T$ would not help in convergence.
Empirically, using the MNIST data set, the proposed algorithm was compared with FedSGD, local training and FedAvg, and was shown to outperform these algorithms.
\appendices
\bibliographystyle{IEEEtran}
\bibliography{IEEEabrv}
%\appendix 

% \clearpage
\begin{appendices}
%\section{Learning Guarantees}
\section{\centerline{\textsc{Proof Of Theorem 1}}}
\label{problem_formulation}
Recall that $\displaystyle S_{k}:=\{(\bm x_{k1}, y_{k1}),(\bm x_{k2},y_{k2}),\ldots,(\bm x_{k n_k},y_{k n_k})\} $ is the set of data samples present at device $k$. Each device $k$ is assumed to have a neural network with weights $\bm w_k \in \mathbb{R}^{d}, k=1,2,\ldots N.$ Let $ W: = \{\bm w_1, \bm w_2, \ldots \bm w_N\} \subseteq \mathcal{R}^{dXN}$ denote the set of neural networks. The neural network weights $\bm w_{k}$ at the device $k$ is optimized  with respect to the loss function $\mathcal{L}_{k}(\bm w_j): =\underset{ (\bm x_k, y_k) \sim \mathcal{D}_k}{\mathbb{E}}\left[l \left(h_{\bm w_j}\left(\bm{x}\right), y\right)\right]$. The weighted sum of the average losses of all the devices in the network is given by
\begin{equation*}
\begin{aligned}
\Phi_{W,\bm{\lambda}} =& \sum_{k=1}^{N}\lambda_k \mathcal{L}_{k}(\bm w_k)
= \sum_{k,j=1}^{N} \lambda_{k}  \alpha_{k j}  \mathcal{L}_{j}(\bm w_k) + \sum_{k,j=1}^{N}\lambda_{k} \alpha_{k j}\Big[ \mathcal{L}_{k}(\bm w_k)- \mathcal{L}_{j}(\bm w_k)\Big],
\end{aligned}
\end{equation*}
where the above is obtained by using the facts that  $\sum_{j=1}^{N} \alpha_{kj}=1$ and $\sum_{k=1}^{N} \lambda_{k}=1$.
Now using the definitions of $\Phi_{ W,\bm{\lambda},\bm{\alpha}} := \sum_{k,j=1}^{N} \lambda_{k} \alpha_{k j}\mathcal{L}_{j}(\bm w_k)$ and the discrepancy between two devices $k$ and $j$ as $\displaystyle d_{kj} := \sup_{\bm{w}}\left|\mathcal{L}_{k}(\bm w)- \mathcal{L}_{j}(\bm w) \right| $, the above can be upper bounded by
\begin{equation}
% \begin{aligned}
\Phi_{ W,\bm{\lambda}} \leq \Phi_{ W,\bm{\lambda},\bm{\alpha}} + \sum_{k,j=1}^{N} \lambda_{k} \alpha_{k j} d_{k j}.
\label{main_bound_eqn}
\end{equation}
Recall that $\hat{\Phi}_{ W,\bm{\lambda},\bm{\alpha}}(\bm S) = \sum_{k=1}^{N} \lambda_k \sum_{j=1}^{N} \alpha_{kj}\hat{\mathcal{L}}(\bm w_k, S_{j}) $ is an estimate of ${\Phi}_{W,\bm{\lambda},\bm{\alpha}}$ 
Towards relating $\hat{\Phi}_{W,\bm{\lambda},\bm{\alpha}}(\bm S)$ to $\Phi_{W,\bm{\lambda},\bm{\alpha}}$, define $\psi\left(\bm S\right) := \sup_{\bm w\in W} \left(\Phi_{W,\bm{\lambda},\bm{\alpha}}-\hat{\Phi}_{W,\bm{\lambda},\bm{\alpha}}(\bm S)\right)$. Consider two set of samples $ \bm S^\prime : = \left\{ S_{1}^\prime, \ldots  S_{N}^\prime \right\}$ and $\bm S=\left\{ S_{1}, \ldots  S_{N} \right\}$ that differ only in a single element, say $(\bm x^\prime_{ki}, y^\prime_{ki}) \in S^\prime _k$ and $(\bm x_{ki}, y_{ki}) \in S _k$ where $k\in \{1\dots N\}$ and $i\in \{1\dots n_k\}$. In order to apply McDiarmid's inequality (see \cite{40597}), one needs to bound the following difference 
\begin{equation*}
\begin{aligned}
\psi\left(\bm S^{\prime}\right)-\psi(\bm S) &=\sup_{\bm w \in W}\left(\Phi_{ W,\bm{\lambda},\bm{\alpha}}-\hat{\Phi}_{ W,\bm{\lambda},\bm{\alpha}}(\bm S^{\prime})\right)-\sup _{\bm w \in W}\left(\Phi_{ W,\bm{\lambda},\bm{\alpha}}-\hat{\Phi}_{ W,\bm{\lambda},\bm{\alpha}}(\bm S)\right) \nonumber\\
& \stackrel{\text{(a)}}{\leq} \sup_{\bm w \in W}\left[\left(\Phi_{ W,\bm{\lambda},\bm{\alpha}}-\hat{\Phi}_{W,\bm{\lambda},\bm \alpha}(\bm S^{\prime})\right)-\left(\Phi_{W,\bm{\lambda},\bm \alpha}-\hat{\Phi}_{W,\bm{\lambda},\bm{\alpha}}(\bm S)\right)\right] \nonumber\\
& {\leq} \sup_{\bm w \in W}\left(\hat{\Phi}_{ W,\bm{\lambda},\bm{\alpha}}(\bm S) - \hat{\Phi}_{ W,\bm{\lambda},\bm{\alpha}}(\bm S^{\prime})\right)\\
& =\sup_{\bm w \in W}\left(\sum_{k,j=1}^{N} \lambda_{k}\alpha_{k j}\left[\hat{\mathcal{L}}(\bm w_k, S_j)- \hat{\mathcal{L}}(\bm w_k, S_j^\prime)\right]\right)   \stackrel{\text{(b)}}\leq \sum_{k=1}^{N}\frac{\lambda_{k}}{n_j}\alpha_{k j}M, %it was n_j and \alpha_{kj}
\label{eq:primarybound}
\end{aligned}
\end{equation*}
where $(a)$ follows from the property of supremum, and $(b)$ follows from the facts that the loss is assumed to be bounded, i.e., $\hat{\mathcal{L}}(\bm w_k, S_j) < M<\infty$, and $(\bm x_{ji}, y_{ji})$ and $(\bm x^\prime_{ji}, y^\prime_{ji})$ differ in only one index $i$. Using McDiarmid's inequality for some $\delta > 0 $ and  $\bm w \in W,$ it is easy to see that the following holds with a probability of at-least $ 1-\delta $ (see \cite{40597})
\begin{equation*}
\begin{aligned}
\Phi_{W,\bm{\lambda},\bm{\alpha}} & \leq \hat{\Phi}_{W,\bm{\lambda},\bm{\alpha}}(\bm S)+\underset{\bm S}{\mathbb{E}}[\psi(\bm S)]+ M\sqrt{ \frac{1}{2} \sum_{j=1}^{N}\left(\sum_{k=1}^{N}\frac{\lambda_{k}\alpha_{k j}}{n_{j}}\right)^{2}\log\left(\frac{1}{\delta}\right)}. \nonumber
\end{aligned}
\end{equation*}
Let $\bm \Lambda_\epsilon$ be an $\epsilon$-cover of $  \bm \Lambda \subseteq \R^n$. By the definition of $\epsilon$-cover (see definition \ref{min_epsilon_cover}), for any $\bm \lambda\in\bm \Lambda, \exists \;\bm \lambda_{\epsilon} \in \bm \Lambda_\epsilon$ such that
$\hat\Phi_{W,\bm{\lambda},\bm \alpha}(\bm S) \leq \hat\Phi_{W,\bm \lambda_{\epsilon},\bm \alpha} (\bm S) + \displaystyle MN\epsilon $. Using this in \eqref{eq:primarybound} with the union bound, the following holds with a probability of at-least $1-\delta$  (see \cite{mohri2019agnostic})
\begin{equation}
\begin{aligned}
\Phi_{W,\bm{\lambda},\bm{\alpha}} \leq \hat{\Phi}_{W,\bm{\lambda}_{\epsilon},\bm{\alpha}} (\bm S)+\underset{\bm S}{\mathbb{E}}[\psi(\bm S)] + MN\epsilon +
M\sqrt{\frac{1}{2} \sum_{j=1}^{N}\left(\sum_{k=1}^{N}\frac{\lambda_{k}\alpha_{k j}}{n_{j}}\right)^{2}\log\left(\frac{\mid\bm\Lambda_{\epsilon}\mid}{\delta}\right)}. 
\label{main_equation}
\end{aligned}
\end{equation}
\vspace{-5mm}
Now, it remains to bound the term $\underset{\bm S}{\mathbb{E}}[\psi(\bm S)]$
%\vspace{0.5mm}
\begin{equation*}
    \begin{aligned}
    \underset{\bm S}{\mathbb{E}}[\psi(\bm S)] &= \underset{\bm S}{\mathbb{E}}\left[\sup _{\bm w\in W}\left(\Phi_{ W,\bm{\lambda},\bm{\alpha}}-\hat{\Phi}_{W,\bm{\lambda},\bm{\alpha}}(\bm S)\right)\right] = \underset{\bm S}{\mathbb{E}}\left[\sup _{\bm w\in W} \underset{\bm S^\prime}{\mathbb{E}}\left(\hat \Phi_{W,\bm{\lambda},\bm{\alpha}}(\bm S^\prime )-\hat{\Phi}_{W,\bm{\lambda},\bm{\alpha}}(\bm S)\right)\right]\\
    & \stackrel{\text{(a)}}\leq \underset{\bm S, \bm S^{\prime}}{\mathbb{E}}\left[\sup _{\bm w\in W} \left(\hat \Phi_{W,\bm{\lambda},\bm{\alpha}}(\bm S^\prime )-\hat{\Phi}_{W,\bm{\lambda},\bm{\alpha}}(\bm S)\right)\right]\\
    &=\underset{\bm S,\bm S^\prime}{\mathbb{E}}\left[\sup _{\bm w\in W}\left(  \sum_{k,j,i=1}^{N,N,n_j} \frac{\lambda_{k} \alpha_{k j} }{n_{j}} \Big[ l(h_{\bm w_{k}}(\bm x_{ji}^\prime), y_{ji}^{\prime})-l(h_{\bm w_{k}}(\bm x_{ji}), y_{ji})\Big]\right)\right],\\
%     \end{aligned}
% \end{equation*}
% \begin{equation*}
%     \begin{aligned}
    % &=\underset{\bm S,\bm S^\prime}{\mathbb{E}}\left[\sup _{\bm w\in W}\left(  
    % \sum_{k,j,i=1}^{N,N,n_j} \frac{\lambda_{k} \alpha_{k j} }{n_{j}} \Big[ l(h_{\bm w_{k}}(\bm x_{ji}^\prime), y_{ji}^{\prime})-l(h_{\bm w_{k}}(\bm x_{ji}), y_{ji})\Big]\right)\right],
    \end{aligned}
\end{equation*}
% \begin{eqnarray}
% \end{eqnarray}
\vspace{-3mm}
%Check this part
where $(a)$ follows from the Jensen's inequality. Since $\left( l(h_{\bm w_{k}}(\bm x_{ji}^\prime), y_{ji}^{\prime})-l(h_{\bm w_{k}}(\bm x_{ji}), y_{ji})\right)
$ and 
\vspace{-3mm}
$\left( l(h_{\bm w_{k}}(\bm x_{ji}), y_{ji})-l(h_{\bm w_{k}}(\bm x_{ji}^\prime), y_{ji}^\prime)\right)
$ have the same distribution, the above can be written as 
\begin{equation*}
\begin{aligned} 
\underset{\bm S}{\mathbb{E}}[\psi(\bm S)] = &\underset{\bm S,\bm S^\prime,\bm \sigma}{\mathbb{E}}\left[\sup _{\bm w\in W}\left(
\sum_{k,j,i=1}^{N,N,n_j} \frac{\sigma_{kji}\lambda_{k} \alpha_{k j} }{n_{j}} \left( l(h_{\bm w_{k}}(\bm x_{ji}^\prime), y_{ji}^{\prime})-l(h_{\bm w_{k}}(\bm x_{ji}), y_{ji})\right)
\right)\right]\\
\stackrel{\text{(a)}} \leq & \underset{\bm S^\prime,\bm \sigma}{\mathbb{E}}\left[
\sup_{\substack{\bm w\in W\\\alpha\in\Delta_N}}
% \sup _{\bm w\in W}
\left(   
\sum_{k,j,i=1}^{N,N,n_j} \mathcal(B)^{'}_{k,j,i} \right)\right]
+\underset{\bm S,\bm \sigma}{\mathbb{E}}\left[
% \sup _{\bm w\in W}
\sup_{\substack{\bm w\in W\\\alpha\in\Delta_N}}
\left(-\sum_{k,j,i=1}^{N,N,n_j}\mathcal(B)_{kji}\right)\right]= 2 \mathcal R_{\bm \Lambda}\left(W \right), 
\end{aligned}
\end{equation*}
{where the Rademacher random} variable $\sigma_{kj,i} \stackrel{\text{iid}} \sim \{-1,1\} $, $\mathcal(B)^{'}_{k,j,i}:= \frac{\sigma_{kji}\lambda_{k} \alpha_{k j}}{n_{j}}  l(h_{\bm w_{k}}(\bm x_{ji}^\prime), y_{ji}^{\prime})$, and $\mathcal(B)_{k,j,i}:=\frac{\sigma_{kji}\lambda_{k} \alpha_{k j}}{n_{j}}  l(h_{\bm w_{k}}(\bm x_{ji}), y_{ji})$. In the above, $(a)$ follows from the fact that $-\sigma_{kji}$ and $\sigma_{kji}$ have the same distribution and the last inequality above follows from the definition of minimax weighted Rademacher complexity, stated in Definition \eqref{rademach}. Using the result above, i.e., $\underset{\bm S}{\mathbb{E}}[\psi(\bm S)] \leq 2 \mathcal R_{\bm \Lambda}\left(W \right)$ in (\ref{main_equation}) results in 
\vspace{-5mm}
\begin{equation}
\begin{aligned}
\Phi_{W,\bm{\lambda},\bm{\alpha}} & \leq \hat{\Phi}_{W,\bm{\lambda},\bm{\alpha}}(\bm S)+  2\mathcal R_{\bm \Lambda}\left(W \right) + MN\epsilon + M\sqrt{\frac{1}{2} \sum_{j=1}^{N}\left(\sum_{k=1}^{N}\frac{\lambda_{k}\alpha_{k j}}{n_{j}}\right)^{2}\log\left(\frac{|\bm \Lambda_{\epsilon}|}{\delta}\right)}. 
\label{loss_bound}
\end{aligned}
\end{equation}
Substituting (\ref{loss_bound}) in (\ref{main_bound_eqn}) results in
\vspace{-5mm}
\begin{equation}
\begin{aligned}
\Phi_{W,\bm{\lambda}} & \leq \hat{\Phi}_{W,\bm{\lambda},\bm{\alpha}}(\bm S)+  2\mathcal R_{\bm \Lambda}\left(W \right) + MN{\epsilon} +
M\sqrt{\frac{N}{2}\log\left(\frac{\mid \bm \Lambda_{ \epsilon}\mid}{\delta}\right) \sum_{k,j=1}^{N}\left(\frac{\lambda_{k}\alpha_{k j}}{n_{j}}\right)^{2}} +
\sum_{k,j=1}^{N}\lambda_{k}\alpha_{kj}d_{kj}, \nonumber
\end{aligned}
\end{equation}
where the term in the square root is upper bounded using the property of norms on vectors in $\R^n$; $\parallel\cdot\parallel_1\leq\sqrt n \parallel\cdot \parallel_2$. This completes the proof.
\qed
%\subsection{Proof of Lemma 1}
\vspace{-5mm}
\section{\centerline{\textsc{Proof of Proposition 1}}}
\label{Validating}
The objective here is to show that the following is a Lipschitz continuous function in $\alpha_{kj}$
\vspace{-4mm}
\begin{equation*}
\begin{aligned} 
\texttt{Reg}(\lambda_k, \bm{\alpha_k})= \frac{M}{N}\sqrt{\frac{N}{2}\log\left(\frac{\mid \bm \Lambda_{ \epsilon}\mid}{\delta}\right) \sum_{j=1}^{N}\sum_{k=1}^{N}\left(\frac{\lambda_{k}\alpha_{k j}}{n_{j}}\right)^{2}}. \nonumber
\label{eqn:square_root}
\end{aligned}
\end{equation*}
Using the mean value theorem, it is sufficient to prove that the norm of the gradient is finite, i.e., each component of $\nabla_{\bm \alpha_k} \texttt{Reg}(\lambda_k, \bm{\alpha_k})$ is finite. The partial derivative of $\texttt{Reg}(\lambda_k, \bm{\alpha_k})$ with respect to $\alpha_{lm}$ for any $l$ and $m$ is given by  
\vspace{-5mm}
\begin{equation}
\begin{aligned}
\frac{\partial \; {\texttt{Reg}(\lambda_k, \bm{\alpha_k})}}{\partial \alpha_{lm}}  =
\frac{M\sqrt{\frac{N}{2}\displaystyle\log\left(\frac{\mid \bm \Lambda_{\epsilon}\mid}{\delta}\right)}\frac{\left(\displaystyle\lambda_{k}^2\alpha_{k m}\right)}{n_m^2}}{N\displaystyle \sqrt{\left(\displaystyle\frac{1}{2} 
\displaystyle \sum_{k,j=1}^{N}\left(\frac{\lambda_{k}\alpha_{kj}}{n_{j}}\right)^{2}\right)}}.
% \frac{\frac{M}{N}\sqrt{\frac{N}{2}\log\left(\frac{\mid \bm \Lambda_{\bm \epsilon}\mid}{\delta}\right)}\frac{\lambda_l}{{n_{m}}}}{\sqrt{\displaystyle \left(1 + \frac{ \sum_{\substack{j=1 \\j \neq m}}^{N}\sum_{\substack{k=1 \\k \neq l}}^{N}\left(\frac{\displaystyle \lambda_{k}\alpha_{kj}}{n_{j}}\right)^{2}}{\scaleto{\left(\frac{\lambda_{l}\alpha_{lm}}{n_{m}}\right)^2}{20pt}}\right)}}. \nonumber
\end{aligned}
\end{equation}
Using the fact that $\displaystyle \frac{1}{\sqrt{1+x}}\leq 1$ for $x\geq0$, 
$
\displaystyle \frac{\partial \; {\texttt{Reg}(\lambda_k, \bm{\alpha_k})}}{\partial \alpha_{lm}} \leq \frac{M}{N}\sqrt{\frac{N}{2}\log\left(\frac{\mid \bm \Lambda_{\epsilon}\mid}{\delta}\right)} \frac{\lambda_{l}}{n_{m}} < \infty, \nonumber
$
% the above can be bounded by
% \begin{align}
% \frac{\partial \; {\texttt{Reg}(\lambda_k, \bm{\alpha_k})}}{\partial \alpha_{lm}} \leq \frac{M}{N}\sqrt{\frac{N}{2}\log\left(\frac{\mid \bm \Lambda_{\epsilon}\mid}{\delta}\right)} \frac{\lambda_{l}}{n_{m}} < \infty, \nonumber
% \end{align}
for all $l$ and $m$. Now, recalling that $\lambda_k \in [0,1]$ and $n_{m} \geq 1$, it is easy to see that the $L_2$-norm of $\nabla_{\bm \alpha_k} \texttt{Reg}(\lambda_k, \bm{\alpha_k})$ is given by
\begin{align*}
\Vert \nabla_{\bm \alpha_k} \texttt{Reg}(\lambda_k, \bm{\alpha_k}) \Vert_{2} \leq \sqrt{\frac{N}{2}\log\left(\frac{\mid \bm \Lambda_{\epsilon}\mid}{\delta}\right) \frac{M^{2}\lambda_{k}}{N^2} \sum_{m=1}^{N}\frac{1}{n_{m}^{2}}} \leq {\beta^{\prime}}: = \frac{M}{\sqrt{2N}}\sqrt{\log\left(\frac{\mid \bm \Lambda_{\epsilon}\mid}{\delta}\right)} < \infty,
\end{align*}
where $\beta^{\prime}$ is the Lipschitz constant. This completes the proof. \qed
% \begin{align}
% \Vert \nabla_{\bm \alpha_k} \texttt{Reg}(\lambda_k, \bm{\alpha_k}) \Vert_2 \leq {\beta^{\prime}}: = \frac{M}{\sqrt{2N}}\sqrt{\log\left(\frac{\mid \bm \Lambda_{\epsilon}\mid}{\delta}\right)} < \infty, \nonumber
% \end{align}

% Note that the Lipschitz `constant of the objective function $\Psi_{W, \bm \lambda, \bm \alpha}$ 
\vspace{-3mm}
\section{\centerline{\textsc{Proof Of Proposition 2}}}
\label{Prop2}
Consider the objective function $\Psi_{\bm w_k, \lambda_k, \bm \alpha_k}$ in ({\ref{MtFEEL_k}}). To prove $ \Psi_{\bm w_k, \lambda_k, \bm \alpha_k}$ is Lipschitz in $\bm \alpha_k$, it suffices to prove 
% that $L_2$ norm of the gradient in $\bm \alpha_k$ is bounded, i.e.,
$\Vert \nabla_{\bm \alpha_k} \Psi_{\bm w_k,  \lambda_k, \bm \alpha} \Vert_2 < \infty $. Taking the partial derivative with respect to $\bm \alpha_{km}$
\begin{equation*}
\begin{aligned}
\frac{\partial \; \Psi_{\bm w_k, \lambda_k, \bm \alpha_k}}{\partial \alpha_{km}} &= \frac{\partial\lambda_k\SUM{N}{j=1}\alpha_{kj}\mathcal{L}_j(\bm w_k)}{\partial \; \alpha_{km}} + \frac{\partial \; \gamma_{k}\parallel \bm w_k \parallel_2}{\partial \alpha_{km}} + \frac{\partial \; \texttt{Reg}({\lambda_k}, \bm{\alpha_k})}{\partial \alpha_{km}} + \frac{\partial \;\lambda_{k} \sum_{j=1}^{N}\alpha_{kj}{ d}_{kj}}{\partial \alpha_{km}}\\
& = \lambda_k\mathcal{L}_m(\bm w_k) + \frac{\partial \; \texttt{Reg}({\lambda_k}, \bm{\alpha_k})}{\partial \alpha_{km}} + \lambda_k { d}_{km} \stackrel{\text{(a)}}{\leq}\lambda_kM + \frac{\partial \; \texttt{Reg}({\lambda_k}, \bm{\alpha_k})}{\partial \alpha_{km}} + \lambda_{k}M,
\end{aligned}
\end{equation*}
where $(a)$ follows by upper bounding the loss, i.e., $\mathcal{L}_m(\bm w_k)$ and discrepancy, i.e., $d_{km}$ by $M$ for all $k$ and $m$. Now, it is easy to see that the $L_2$-norm of $\nabla_{\bm \alpha_k} \Psi_{\bm w_k, \lambda_k, \bm \alpha_k}$ is given by
$$
\Vert \nabla_{\bm \alpha_k} \Psi_{\bm w_k, \lambda_k, \bm \alpha_k}\Vert_2 \leq 
\lambda_kM + \Vert \nabla_{\bm \alpha_k} \texttt{Reg}(\lambda_k, \bm{\alpha_k}) \Vert_2 + \lambda_{k}M \stackrel{\text{(b)}}{=} \beta^{\prime} + 2 \lambda_kM < \infty,
$$
\vspace{-2.3mm}
where ${\beta^{\prime}}: = \frac{M}{\sqrt{2N}}\sqrt{\log\left(\frac{\mid \bm \Lambda_{\epsilon}\mid}{\delta}\right)}$, and $(b)$ follows from Proposition 1. Next it remains to prove that $\Psi_{\bm w_k, \lambda_k, \bm \alpha_k}$ is Lipschitz in $\bm w_{k}$. The gradient with respect to $\bm w_k$ is given by \footnote{For simplicity $\gamma_k$ is assumed to be $0$.}
% \begin{equation}
$
\nabla_{\bm w_k} \Psi_{\bm w_k, \lambda_k, \bm \alpha_k}  = \lambda_k\SUM{N}{j=1}\alpha_{kj} \texttt{sign}({\bm g}_{kj}) 
$
% \end{equation}
and the $L_2$-norm of $\nabla_{\bm w_k} \Psi_{\bm w_k, \lambda_k, \bm \alpha_k}$ is
$
\Vert \nabla_{\bm w_k} \Psi_{\bm w_k, \lambda_k, \bm \alpha_k} \Vert_2 \stackrel{\text{(c)}}{\leq} \lambda_kd.
$
Here, $(c)$ follows from the fact that $\SUM{N}{j=1}\alpha_{kj} = 1$ and $\texttt{sign}({\bm g}_{kj}) \leq 1$ for all $k$ and $j$.
The Lipschitz constant of $ \Psi_{\bm w_k, \lambda_k, \bm \alpha_k} $ will thus be $U: = \lambda_kd + \beta^{\prime} + 2 \lambda_kM$.
\vspace{-5mm}
\section{\centerline{\textsc{Proof Of Theorem 2}}}
\label{thrm2}
% In order to prove convergence of the algorithm, it suffices to prove that $\bm w_k$ converges for all $k=1,2, \ldots, N$. Equivalently, it is enough to show that the gradient of
% \begin{equation}
% \Psi_{\bm w_k, \lambda_k, \bm \alpha_k} = \Phi_{\bm w_k,{\lambda_k},\bm{\alpha_k}}(\bm S) + \gamma_{k}\parallel \bm w_k \parallel_2 + \texttt{Reg}({\lambda_k}, \bm{\alpha_k}) + \lambda_{k} \sum_{j=1}^{N}\alpha_{kj}{ d}_{kj}, 
% \end{equation}
% which corresponds to the $k$-th component of $\Psi_{W,\bm \lambda, \bm \alpha}$, converges. Further,  \\
% $\Phi_{\bm w_k,{\lambda_k},\bm{\alpha_k}}(\bm S): = \lambda_k \sum_{j=1}^{N} \alpha_{kj} \mathcal{L}_j(\bm w_k),$ and 
% $\texttt{Reg}({\lambda_k}, \bm{\alpha_k}): = \frac{\texttt{Reg}(\bm{\lambda}, \bm{\alpha})}{N}$.
% $\texttt{Reg}(\bm{\lambda_k}, \bm{\alpha_k}): = \frac{M}{N}\sqrt{\frac{1}{2} \sum_{j=1}^{N}\left(\sum_{k=1}^{N}\frac{\lambda_{k}\alpha_{k j}}{n_{j}}\right)^{2}\log\left(\frac{\mid\Lambda_{\bm \epsilon}\mid}{\delta}\right)}.$ 
% It is enough to show the convergence for each device $k$, in this objective as a linear combination of the individual components will also converge, hence defining for each device $k=1,2\ldots N$
%\label{convergence proof}
Consider the $\beta$-smoothness assumption (see \ref{assumption 2}) of $ \Psi_{\bm w_k, \lambda_k, \bm \alpha_k}$ with respect to $\bm w_k$ for all $k= 1,2, \ldots, N $ in \eqref{MtFEEL_k}
\begin{equation}
\begin{aligned}
\Psi_{\bm w_k^{t+1},  \lambda_k, \bm \alpha_k^{t+1}}
-\Psi_{\bm w_k^{t}, \lambda_k, \bm \alpha_k^{t+1}}&
\leq \langle \nabla_{\bm w_k} \Psi_{\bm w_k^{t},  \lambda_k, \bm \alpha_k^{t+1}}, \bm w_{k}^{t+1}-\bm w_{k}^{t}\rangle +
\sum_{i=1}^{d} \frac{L_{i}}{2}\left(\bm w_{k}^{t+1}-\bm w_{k}^{t}\right)_{i}^{2}, \nonumber
% \Phi_{(\bm w_k^{t+1}, \bm \alpha_{k}^{t+1}, \bm\lambda)}-\Phi_{\bm w_k^{t}, \bm \alpha_{k}^{t+1}, \bm\lambda} &
% \leq \langle \nabla_{\bm w_k} \Phi_{\bm w_k^{t}, \bm \alpha_{k}^{t+1}, \bm\lambda}, \bm w_{k}^{t+1}-\bm w_{k}^{t}\rangle +
% \sum_{i=1}^{d} \frac{L_{i}}{2}\left(\bm w_{k}^{t+1}-\bm w_{k}^{t}\right)_{i}^{2}, \nonumber
\label{smoothness_ass}
\end{aligned}
\end{equation}
where $(\bm x)_i$ is the $i^{th}$ component of the vector $\bm x \in \mathbb{R}^d$. Define $\Delta_{\bm w_{k}} \Psi_{\bm w_k^{t+1}, \lambda_k, \bm \alpha_k^{t+1}} := \Psi_{\bm w_k^{t+1}, \lambda_k, \bm \alpha_k^{t+1}}-\Psi_{\bm w_k^{t}, \lambda_k, \bm \alpha_k^{t+1}}$.  Now, using $\bm w_{k}^{t+1}- \bm w_{k}^{t} = -\eta^{t}\nabla_{\bm w_{k}, \operatorname{sign}}\hat \Psi_{\bm w_k^{t}, \lambda_k, \bm \alpha_k^{t+1}}$ from step 7 of the \textbf{Algorithm $2$}, the above becomes
\vspace{-5mm}
\begin{align}
\label{eq:Lipschitz}
\Delta_{\bm w_{k}} \Psi_{\bm w_k^{t+1}, \lambda_k, \bm \alpha_k^{t+1}} & \leq -\eta^{t}\nabla_{\bm w_{k}} \Psi_{{\bm w_k^{t}, \lambda_k, \bm \alpha_k^{t+1}}}^ {T}\nabla_{\bm w_{k}, \operatorname{sign}}\hat\Psi_{\bm w_k^{t}, \lambda_k, \bm \alpha_k^{t+1}} +\sum_{i=1}^{d}\frac{L_{i}}{2}\left(-\eta^{t}\nabla_{\bm w_{k}, \operatorname{sign}}\hat\Psi_{\bm w_k^{t}, \lambda_k, \bm \alpha_k^{t+1}}\right)_{i}^{2}. \nonumber
\end{align}
% \vspace{-5mm}
Substituting for the true gradient $\nabla_{\bm w_{k}} \Psi_{\bm w_k^{t},  \lambda_k, \bm \alpha_k^{t+1}} = \lambda_{k} \sum_{m=1}^{N} \alpha_{k m}^{t+1} \bm g_{km }^t ,$ and its estimate \\$\nabla_{\bm w_{k},\operatorname{sign}} \hat\Psi_{\bm w_k^{t}, \lambda_k, \bm \alpha_k^{t+1}} =\lambda_{k} \sum_{j=1}^{N} \alpha_{k j}^{t+1} \texttt{sign}\left(\hat{\bm g}_{k j}^t\right) $, and after some algebraic manipulations, the above becomes
% \begin{equation}
% \begin{aligned}
% & = -\eta^{t}\left(\lambda_{k} \sum_{m=1}^{N} \alpha_{k m}^{t+1} (\bm g_{km }^t)^{T}\right) 
% \left( \lambda_{k} \sum_{j=1}^{N} \alpha_{k j}^{t+1} \operatorname{sign}\left(\hat{\bm g}_{k j}^t\right)\right)+ 
% \frac{(\eta^{t})^{2}}{2}\sum_{i=1}^{d}\frac{L_{i}}{2}\left(\lambda_{k} \sum_{j=1}^{N} \alpha_{k j}^{t+1} \operatorname{sign}\left(\hat{\bm g}_{k j}^t\right)\right)_{i}^{2}. \nonumber\\
% \end{aligned}
% \end{equation}
% After some algebraic manipulations, the above becomes
\vspace{-5mm}
\begin{equation}
\begin{aligned}
\Delta_{\bm w_{k}} \Psi_{\bm w_k^{t+1}, \lambda_k, \bm \alpha_k^{t+1}} & \leq -\eta^{t}\left(\lambda_{k}^{2} \sum_{m=1}^{N}  \alpha_{k m}^{t+1}{\parallel \bm g_{k m}^t\parallel}_{1} \right) +
2 \eta^{t}\Bigg(\lambda_{k}^{2}\sum_{m=1}^{N} \sum_{i=1}^{d} {\alpha_{km}^{t+1}}| g_{km,i}^t|\mathcal{F}_{k,i}^t \Bigg)\\
&+ \frac{(\eta^{t})^{2}}{2}\sum_{i=1}^{d}L_{i}\left(\lambda_{k} \sum_{j=1}^{N} \alpha_{k j}^{t+1} \operatorname{sign}\left(\hat{\bm g}_{k j}^t\right)\right)_{i}^{2}, \nonumber
\label{upper_bound_sign}
\end{aligned}
\vspace{-7mm}
\end{equation}
where $\mathcal{F}_{k,i}^t: = \sum_{j=1}^{N} \alpha_{kj}^{t+1}\mathbbm{1}\left[\texttt{sign}\left(\hat{g}_{kj, i}^t\right) \neq \texttt{sign}\left( g_{km,i}^t\right)\right]\Bigg)$. 
Using the fact that the $i^{th}$ component of the signed gradient is less than or equal to $1$, i.e., $\texttt{sign}\left(\hat{\bm g}_{k j, i}^t\right) \leq 1$, the above can be further upper bounded as
\vspace{-7mm}
\begin{equation}
\begin{aligned}
\Delta_{\bm w_{k}} \Psi_{\bm w_k^{t+1}, \lambda_k, \bm \alpha_k^{t+1}} & \leq -\eta^{t}\left(\lambda_{k}^{2} \sum_{m=1}^{N}  \alpha_{k m}^{t+1}{\parallel \bm g_{k m}^t\parallel}_{1} \right) +
2 \eta^{t}\Bigg(\lambda_{k}^{2}\sum_{m=1}^{N} \sum_{i=1}^{d} {\alpha_{km}^{t+1}}| g_{km,i}^t|\mathcal{F}_{k,i}^t \Bigg) \\ +& \frac{(\eta^{t})^{2}}{2}\parallel L\parallel_{1}\lambda_{k}^{2} .\nonumber
\end{aligned}
\end{equation}
\vspace{-2mm}
Consider the expected improvement at $t+1$ conditioned on the previous iterate, i.e.,
%\begin{equation}
\begin{align}
&\mathbb{E}\left[\Delta_{\bm w_{k}} \Psi_{\bm w_k^{t+1}, \lambda_k, \bm \alpha_k^{t+1}} \mid \bm w_{k}^{t}\right] \leq -\eta^{t}\left(\lambda_{k}^{2} \sum_{m=1}^{N} \alpha_{km}^{t+1}{\parallel \bm g_{k m}^t\parallel}_{1}\right) + \nonumber\\ 
& 2 \eta^{t}\Bigg(\lambda_{k}^{2}\sum_{m,j=1}^{N}\sum_{i=1}^{d}{ \alpha_{km}}\mid g_{km,i}^t\mid\mathbb{P}\left[\texttt{sign}\left(\hat{g}_{kj, i}^t\right) \neq \texttt{sign}\left( g_{km,i}^t\right)\right]\Bigg)+ 
\frac{(\eta^{t})^2}{2} \parallel L\parallel_{1}\lambda_{k}^{2}.
\label{Improvement_equation}
\end{align}
\vspace{-6mm}
In order to bound the above further, consider the term
\begin{eqnarray}
\mathbb{P}\left[\texttt{sign}\left(\hat{g}_{k j,i}^t\right) \neq \texttt{sign}\left(g_{k m, i}^t\right)\right] & \stackrel{\text{(a)}}{\leq} \mathbb{P}\left[\mid \hat{g}_{kj,i}^t - g_{km, i}^t\mid \geq \mid g_{km,i}^t\mid\right] \nonumber \stackrel{\text{(b)}}{\leq}& \frac{\mathbb{E}\left[\left|\hat{g}_{k j, i}^t- g_{k m, i}^t\right|\right]}{\left|g_{k m, i}^t\right|} \nonumber \\
&\stackrel{(c)}\leq \frac{\sqrt{\mathbb{E}\left[{\left(\hat{g}_{k j, i}^t- g_{k m, i}^t\right)}^2\right]}}{\left|g_{k m, i}^t\right|} \stackrel{(d)}{\leq} \frac{\sigma_{k m,i}^t}{\left|g_{k m, i}^t\right|}.
\label{Probability_gradients}
\end{eqnarray}
In the above, $(a)$ follows from the fact that $\{\texttt{sign}\left(\hat{g}_{k j,i}^t\right) \neq \texttt{sign}\left(g_{k m, i}^t\right)\} \subseteq \{\mid \hat{g}_{kj,i}^t - g_{km, i}^t\mid \geq \mid g_{km,i}^t\mid \}$, $(b)$ follows from the Markov's inequality, $(c)$ is obtained from the Jensen's inequality, and $(d)$ follows since $\hat g_{kj,i}^t$ is an unbiased estimate of $g_{km,i}^t$ and using the definition of variance. The following is obtained by substituting (\ref{Probability_gradients}) in (\ref{Improvement_equation}) and using $\sigma_{km, i}^t \leq \frac{\sigma_{k m,i}}{\sqrt{n_{t}}}$ 
\vspace{-5mm}
\begin{align}
\mathbb{E}\left[\Delta_{\bm w_{k}} \Psi_{\bm w_k^{t+1}, \lambda_k, \bm \alpha_k^{t+1}} \mid \bm w_{k}^{t+1}\right] \leq  -\eta^{t}&\left(\lambda_{k}^{2} \sum_{m=1}^{N}  \alpha_{k m}^{t+1}\left\|\bm g_{k m}^t\right\|_{1}\right) +  2\eta^{t}\left(\lambda_{k}^{2} \sum_{m=1}^{N}\frac{{ \alpha_{km}^{t+1}}\parallel \bm \sigma_{k m}\parallel_{1}}{\sqrt{n_{t}}}\right)  \nonumber\\
+ &\frac{(\eta^{t})^2}{2} \parallel L\parallel_{1}\lambda_{k}^{2}. \nonumber 
\end{align}
Since $\alpha_{km} \leq 1$ for all $k$ and $m$, the above can be further bounded as
\begin{align}
\leq -\eta^{t}\left(\lambda_{k}^{2}\sum_{m=1}^{N}  \alpha_{k m}^{t+1}\left\|\bm g_{k m}^t\right\|_{1}\right) + \left(2\eta^{t}\lambda_{k}^{2} \sum_{m=1}^{N}\frac{\parallel \bm \sigma_{k m}\parallel_{1}}{\sqrt{n_{t}}}\right) + \frac{(\eta^{t})^2}{2} \parallel L\parallel_{1}\lambda_{k}^{2}. 
\label{weights_eqn}
\end{align}
Now, bounding the difference of the objective function when $\bm w_{k}^t$ is fixed while $\bm \alpha_{k}^t$ is a variable
\begin{align}
\Psi_{\bm w_k^{t}, \lambda_k, \bm \alpha_k^{t+1}}-\Psi_{\bm w_k^{t}, \lambda_k, \bm \alpha_k^{t}} & \leq\left\langle\nabla_{\bm \alpha_{k}} \Psi_{\bm w_k^{t}, \lambda_k, \bm \alpha_k^{t}} , \bm \alpha_{k}^{t+1}-\bm \alpha_{k}^{t}\right\rangle+\frac{\beta}{2}{\parallel \bm \alpha_{k}^{t+1}-\bm \alpha_{k}^t\parallel}_{2}^{2}, \nonumber
\end{align}
where $\beta : = M\Big(\sqrt{\frac{1}{2N}\log\left(\frac{\mid \bm \Lambda_{\epsilon}\mid}{\delta}\right)} + 2\lambda_k \Big)$ is as defined in Lemma 2. Substituting $\bm \alpha_{k}^{t+1}-\bm \alpha_{k}^{t}$ = $-\mu^{t}\nabla_{\mathcal K,\bm \alpha_{k}} \hat \Psi_{\bm w_k^{t}, \lambda_k, \bm \alpha_k^{t}}$ from step 8 of the \textbf{Algorithm $2$}, where $\nabla_{\mathcal K,\bm \alpha_{k}} \hat \Psi_{\bm w_k^{t}, \lambda_k, \bm \alpha_k^{t+1}}$ denotes the projected gradient with respect to $\bm \alpha_{k}$, the difference $\Delta_{\bm \alpha_{k}}\Psi_{\bm w_k^{t+1}, \lambda_k, \bm \alpha_k^{t+1}}:= \Psi_{\bm w_k^{t}, \lambda_k, \bm \alpha_k^{t+1}}-\Psi_{\bm w_k^{t}, \lambda_k, \bm \alpha_k^{t}} $ becomes
%\begin{align}
$\Delta_{\bm \alpha_{k}}\Psi_{\bm w_k^{t}, \lambda_k, \bm \alpha_k^{t}} =-\mu^{t}\left\langle\nabla_{\bm \alpha_{k}} \Psi_{\bm w_k^{t}, \lambda_k, \bm \alpha_k^{t}}, \nabla_{\mathcal K, \bm \alpha_{k}} \hat \Psi_{\bm w_k^{t}, \lambda_k, \bm \alpha_k^{t}} \right\rangle+\frac{\beta(\mu^{t})^{2}}{2} \left\Vert\nabla_{\mathcal K,\bm \alpha_{k}} \hat \Psi_{\bm w_k^{t}, \lambda_k, \bm \alpha_k^{t}}\right\Vert_{2}^{2}$.
%\label{alpha_smoothness}
%\end{align}
Using $\left\langle\nabla_{\bm \alpha_{k}} \Psi_{\bm w_k^{t}, \lambda_k, \bm \alpha_k^{t}}, \nabla_{\mathcal K, \bm \alpha_{k}} \hat \Psi_{\bm w_k^{t}, \lambda_k, \bm \alpha_k^{t}}\right\rangle \geq \Vert \nabla_{\mathcal K, \bm \alpha_{k}} \hat \Psi_{\bm w_k^{t}, \lambda_k, \bm \alpha_k^{t}} \Vert_{2}^2$ from Lemma 3.2 of \cite{hazan2017efficient}, the above can be further bounded as
%\begin{equation}
%\left\langle\nabla_{\bm \alpha_{k}} \Psi_{\bm w_k^{t}, \lambda_k, \bm \alpha_k^{t}}, \nabla_{\mathcal K, \bm \alpha_{k}} \hat \Psi_{\bm w_k^{t}, \lambda_k, \bm \alpha_k^{t}}\right\rangle \geq \Vert \nabla_{\mathcal K, \bm \alpha_{k}} \hat \Psi_{\bm w_k^{t}, \lambda_k, \bm \alpha_k^{t}} \Vert_{2}^2 
%\end{equation}
\vspace{-5mm}
\begin{equation}
\Delta_{\bm \alpha_{k}} \Psi_{\bm w_k^{t+1}, \lambda_k, \bm \alpha_k^{t}} \leq -\mu^{t}\left\Vert \nabla_{\mathcal K, \bm \alpha_{k}} \hat \Psi_{\bm w_k^{t}, \lambda_k, \bm \alpha_k^{t}}\right\Vert_{2}^{2}+\frac{\beta (\mu^{t})^2}{2}\left\Vert\nabla_{\mathcal K, \bm \alpha_{k}} \hat \Psi_{\bm w_k^{t}, \lambda_k, \bm \alpha_k^{t}}\right\Vert_{2}^{2}. 
\label{eq:alpha_update}
\end{equation}
\vspace{-7mm}
Let $\Psi_k^* := \underset{\bm w_k,\bm \alpha_k}{\min} \Psi_{\bm w_k, \lambda_k \bm \alpha_k}$ such that $\sum_{j=1}^{N} \alpha_{kj} = 1 $. Towards completing the proof, consider
% \vspace{-1mm}
\begin{eqnarray*}
\Psi_{\bm w_k^{0}, \lambda_k, \bm \alpha_k^{0}}- \Psi_{k}^{*}  & \stackrel{\text{(a)}} \geq& \Psi_{\bm w_k^{0}, \lambda_k, \bm \alpha_k^{0}} - \mathbb{E}[\Psi_{\bm w_k^{t+1}, \lambda_k, \bm \alpha_k^{t+1}}] \stackrel{\text{(b)}} = \mathbb{E}\sum_{t=0}^{T-1}[\Psi_{\bm w_k^{t},  \lambda_k, \bm \alpha_k^{t}} - \Psi_{\bm w_k^{t+1},  \lambda_k, \bm \alpha_k^{t+1}}]\\
% \end{eqnarray*}
% \begin{eqnarray*}
& \stackrel{\text{(c)}} =& \mathbb{E}\sum_{t=0}^{T-1}\Big[-\Big(\Psi_{\bm w_k^{t+1}, \lambda_k, \bm \alpha_k^{t+1})}-\Psi_{\bm w_k^{t}, \lambda_k, \bm \alpha_k^{t+1}}\Big) - \Big(\Psi_{\bm w_k^{t}, \lambda_k, \bm \alpha_k^{t+1}}-\Psi_{\bm w_k^{t}, \lambda_k, \bm \alpha_k^{t}}\Big)\Big] \nonumber\\
&=& -\mathbb{E}\sum_{t=0}^{T-1}\Delta_{\bm w_{k}} \Psi_{\bm w_k^{t+1}, \lambda_k, \bm \alpha_k^{t+1}} - \mathbb{E}\sum_{t=0}^{T-1} \Delta_{\bm \alpha_{k}}\Psi_{\bm w_k^{t+1}, \lambda_k, \bm \alpha_k^{t+1}}, \label{eq:telescopic_sum_split}
\end{eqnarray*}
% \vspace{-3mm}
where $(a)$ follows from the optimality of $\Psi_k^*$, $(b)$ follows from the telescopic sum and $(c)$ follows by adding and subtracting $\Psi_{\bm w_k^{t}, \lambda_k, \bm \alpha_k^{t+1}}$.
Substituting (\ref{weights_eqn}) and (\ref{eq:alpha_update}) in (\ref{eq:telescopic_sum_split}) and choosing the learning rates $\eta^{t}=\frac{1}{\sqrt{T}} $ and $ \mu^t=\frac{1}{\sqrt{T}}$, and batch size $ n_{t} = T$ results in \eqref{eq:conv_alg2}. \qed
\section{\centerline{\textsc{Proof Of Theorem \ref{thm: Theorem3}}}}
\label{convergence with distortion proof:expectation}
% From equation (\ref{eq: theorem2}), of \textbf{Theorem} \ref{Theorem 2}, the following holds:
% \begin{equation}
% \begin{aligned}
% \mathbb{E}\Bigg[\frac{1}{T}\sum_{t=0}^{T-1}\left(\lambda_{k}^{2}\sum_{m=1}^{N}  \alpha_{k m}^{t+1}\left\|\bm g_{k m}^t\right\|_{1}\right)\Bigg] + \left(1-\frac{\beta }{2\sqrt{T}}\right) \mathbb{E}\left[\frac{1}{T}\sum_{t=0}^{T-1}\left\|\nabla_{\mathcal K, \bm \alpha_{k}} \Psi_{\bm w_{k}^{t}, \lambda_k, \bm \alpha_{k}^{t}}\right\|_2^{2}\right] \leq \nonumber\\ \frac{1}{\sqrt{T}}\Bigg(2\lambda_{k}^{2} \sum_{m=1}^{N}\parallel \bm \sigma_{k m}\parallel_{1} +  \frac{\parallel L \parallel_{1} \lambda_{k}^{2}}{2} + \Psi_{\bm w_k^{0}, \lambda_k, \bm \alpha_k^{0}} - \Psi_{k}^{*}\Bigg).\nonumber
% \label{eq:thrm4}
% \end{aligned}
% \end{equation}
% The term $\Delta_k^t$ is defined from the left hand side of the above as \[\Delta_k^t:=\Bigg[\left(\lambda_{k}^{2}\sum_{m=1}^{N} \alpha_{k m}^{t+1}\left\|\bm g_{k m}^t\right\|_{1}\right) + \left(1-\frac{\beta }{2\sqrt{T}}\right)\left[\left\|\nabla_{\mathcal K, \bm \alpha_{k}} \Psi_{\bm w_{k}^{t}, \lambda_k, \bm \alpha_{k}^{t}}\right\|_2^{2}\right]\Bigg].\]
% from \textbf{Theorem} \ref{Theorem 2}, 
% if there are $d$ bits of information to be sent across a wireless channel, the average. \E{}{[\Delta_k^t]}$ goes down to $0$ asymptotically with $t$ for all $k$. 
It suffices to show that $\frac{1}{T} \sum_{t=1}^T \mathbb{E} [\Delta_k^t] \to 0$ as $T \to \infty$ in a Rayleigh fading channel. Here,
% The aim is to find the rate at which $\frac{1}{T} \sum_{t=1}^T \mathbb{E} [\Delta_k^t] \to 0$ as $T \to \infty$ in a Rayleigh fading channel. Here,
$\Delta_k^t:=\Bigg[\left(\lambda_{k}^{2}\sum_{m=1}^{N} \alpha_{k m}^{t+1}\left\|\bm g_{k m}^t\right\|_{1}\right) + \left(1-\frac{\beta }{2\sqrt{T}}\right)\left[\left\|\nabla_{\mathcal K, \bm \alpha_{k}} \Psi_{\bm w_{k}^{t}, \lambda_k, \bm \alpha_{k}^{t}}\right\|_2^{2}\right]\Bigg]$. From the total expectation law, it follows that
% $ \Delta_t^k$ is a sequence of random variables, each of which is upper bounded using the Lipshcitz constants from \textbf{Lemma} \ref{lipconstants},
% $$\Delta_T^k\leq U:=\lambda_kL + \left(1-\frac{\beta }{2\sqrt{T}}\right){\beta}$$ 
% for $T\geq \frac{\beta^2}{4}$. Calling the event of an outage $O$, from the total expectation law,
\begin{equation*}
    \begin{aligned}
    \mathbb{E}[\Delta_k^t]=& \mathbb{E}[\Delta_k^t | \mathcal{O}] \mathbb{P}[O]+\mathbb{E}\left[\Delta_k^t | \mathcal{O}^{\complement}\right] \mathbb{P}[\mathcal{O}^{\complement}]
{\leq}  U\mathbb{P}[\mathcal{O}]+\mathbb{E}\left[\Delta_k^t | \mathcal{O}^{\complement}\right] \mathbb{P}[\mathcal{O}^{\complement}],
    \end{aligned}
\end{equation*}
where $\mathcal{O}$ is the outage event. The inequality above follows from the fact that $\Delta_k^t$ is bounded above by the Lipschitz constant $U: = \beta^{\prime} + 2 \lambda_kM + \lambda_kd $ from lemmas \ref{lem:lips_reg} and \ref{lem:lips_psihat}. Now from the definition of outage event, and using the fact that $\mathbb{P}[O^{\complement}] \leq 1$, the above can be bounded as
% , therefore $\mathbb{E}[\Delta_k^t | O]$ is also upper bounded by $U$, 
\begin{equation*}
\begin{aligned}
\mathbb{E}[\Delta_k^t] \leq &
 U \mathbb{P}\Bigg\{d\geq B\ln\left(1+\frac{\mathcal P_k |h_k|^2}{B\sigma^2}\right)\Bigg\} + \mathbb{E}\left[\Delta_k^t | O^{\complement}\right] \nonumber \\ =& U\left(1-\exp\left\{-\displaystyle \frac{(2^{\frac{d}{B}}-1)}{SNR_k}\right\} \right)  + \mathbb{E}\left[\Delta_k^t | O^{\complement}\right].
% \stackrel{\text{(b)}}
% \leq & U \mathbb{P}\Bigg{d\geq B\ln\left(1+\frac{\mathcal P_k |h_k|^2}{B\sigma^2}\right)\Bigg}+\frac{1}{\sqrt{T}}\Bigg(2\lambda_{k}^{2} \sum_{m=1}^{N}\parallel \bm \sigma_{k m}\parallel_{1} +  \frac{\parallel L \parallel_{1} \lambda_{k}^{2}}{2} + \Psi_{\bm w_k^{0}, \lambda_k, \bm \alpha_k^{0}} - \Psi_{k}^{*}\Bigg)\\
    \end{aligned}
\end{equation*}
% \vspace{-5mm}
Using the above, the average of $\Delta_k^t$ is bounded as follows
\begin{equation*}
    \frac{1}{T} \sum_{t=1}^T \mathbb{E}\Delta_k^t \stackrel{\text{(a)}}\leq  U\frac{(\displaystyle 2^{\frac{d}{B}}-1)}{SNR_k}+
 \frac{1}{\sqrt{T}}\Bigg(2\lambda_{k}^{2} \sum_{m=1}^{N}\parallel \bm \sigma_{k m}\parallel_{1} +  \frac{\parallel L \parallel_{1} \lambda_{k}^{2}}{2} + \Psi_{\bm w_k^{0}, \lambda_k, \bm \alpha_k^{0}} - \Psi_{k}^{*}\Bigg),
\end{equation*}
% \vspace{-4mm}
where the above follows from the facts that $1-e^{-x}\leq x$ and that when there is no outage, the analysis is identical to the case in Appendix \ref{thrm2}. 
This completes the proof of the theorem.
\qed

\end{appendices}

% that's all folks
\end{document}